\documentclass[sigconf]{acmart}

\AtBeginDocument{%
  \providecommand\BibTeX{{%
    \normalfont B\kern-0.5em{\scshape i\kern-0.25em b}\kern-0.8em\TeX}}}

\copyrightyear{2021} 
\acmYear{2021} 
\setcopyright{acmcopyright}\acmConference[SIGMOD '21]{Proceedings of the 2021 International Conference on Management of Data}{June 20--25, 2021}{Virtual Event, China}
\acmBooktitle{Proceedings of the 2021 International Conference on Management of Data (SIGMOD '21), June 20--25, 2021, Virtual Event, China}
\acmPrice{15.00}
\acmDOI{10.1145/3448016.3459240}
\acmISBN{978-1-4503-8343-1/21/06}

\usepackage{geometry}
\usepackage{marginnote}

\usepackage{graphicx}
\usepackage{balance}  

\usepackage[utf8]{inputenc} 
\usepackage[T1]{fontenc}    
\usepackage{url}            
\usepackage{booktabs}       
\usepackage{amsfonts}       
\usepackage{nicefrac}       
\usepackage{microtype}      
\usepackage{amsmath}
\usepackage{bm}
\usepackage{graphicx}
\usepackage{algorithm}
\usepackage[noend]{algpseudocode}
\usepackage{xspace}
\usepackage{color,soul}
\usepackage{multirow}
\usepackage{multicol}
\usepackage{caption}
\usepackage{subfig,float}

\usepackage{enumitem}
\setenumerate[1]{itemsep=0pt,partopsep=0pt,parsep=0.1\parskip,topsep=0pt,leftmargin=5pt}
\setitemize[1]{itemsep=0pt,partopsep=0pt,parsep=0.1\parskip,topsep=0pt,leftmargin=5pt}

\newcommand{\system}{\text{LambdaML}\xspace}

\usepackage[most]{tcolorbox}
\tcbset{width=1\columnwidth,boxrule=0pt,colback=lightgray,arc=0pt,auto outer arc,left=0pt,right=0pt,boxsep=2pt}

\definecolor{myred}{rgb}{1.0,0.7,0.8}
\definecolor{mygreen}{RGB}{0,166,0}
\definecolor{lightgreen}{rgb}{0.56, 0.93, 0.56}
\definecolor{myorange}{RGB}{252,107,4}
\definecolor{darkgreen}{RGB}{0,153,102}
\definecolor{lightblue}{rgb}{0.53, 0.81, 0.92}
\definecolor{lightgray}{gray}{0.8}

\newcommand{\remove}[1]{}

\newcommand{\papertitle}{Towards Demystifying Serverless Machine Learning Training}

\begin{document}

\title{\papertitle}

\author{Jiawei Jiang$^{*,\dagger}$, Shaoduo Gan$^{*,\dagger}$, Yue Liu$^{\dagger}$, Fanlin Wang$^{\dagger}$\\
Gustavo Alonso$^{\dagger}$, Ana Klimovic$^{\dagger}$, 
Ankit Singla$^{\dagger}$, Wentao Wu$^{\#}$, Ce Zhang$^{\dagger}$}
\affiliation{
  \institution{$^{\dagger}$Systems Group, ETH Z\"{u}rich~~$^{\#}$Microsoft Research, Redmond \\
  \{jiawei.jiang, sgan, alonso, ana.klimovic, ankit.singla, ce.zhang\}@inf.ethz.ch, \\ 
  \{liuyue, fanwang\}@student.ethz.ch, wentao.wu@microsoft.com}
  \city{}
  \country{}
}


\renewcommand{\shortauthors}{Jiawei Jiang, Shaoduo Gan, Yue Liu, Fanlin Wang, Gustavo Alonso, Ana Klimovic, Ankit Singla, Wentao Wu, and Ce Zhang}

\renewcommand{\authors}{Jiawei Jiang, Shaoduo Gan, Yue Liu, Fanlin Wang, Gustavo Alonso, Ana Klimovic, Ankit Singla, Wentao Wu, Ce Zhang}

\begin{abstract}

The appeal of serverless (FaaS) has triggered a growing interest on how
to use it in data-intensive applications such as ETL, query processing, or machine learning (ML). Several systems exist for training large-scale ML models on top of serverless infrastructures
(e.g., AWS Lambda) but with inconclusive results in terms of their performance and relative advantage over
``serverful'' infrastructures (IaaS). In
this paper we present a systematic, comparative study of distributed ML training over FaaS
and IaaS. We present a
design space covering design choices
such as optimization algorithms and synchronization
protocols, and implement a platform, LambdaML, that enables a fair
comparison between FaaS and IaaS. We present experimental
results using LambdaML, and further develop an analytic model to capture cost/performance
tradeoffs that must be considered when opting for a serverless infrastructure. Our results indicate that ML training pays off in serverless only for models with efficient (i.e., reduced) communication and that quickly converge. In general, FaaS can be much faster but it is never significantly cheaper than IaaS.  

\end{abstract}



\fancyhead{} 


\maketitle

\renewcommand\thefootnote{}\footnote{*Equal contribution.}

\section{Introduction} \label{sec:intro}

Serverless computing has recently emerged as a new type of computation infrastructure. While initially developed for web microservices and IoT applications, recently researchers have
explored the role of serverless computing in data-intensive 
applications, which stimulates intensive interests in the data management community~\cite{pocket-storage,openlambda,serverless-ATC18,SAND-ATC18,serverless-HotEdge19}.
Previous work has shown that adopting a serverless infrastructure for \textit{certain} types of workloads
can significantly lower the cost. 
Example workloads range from ETL~\cite{USETL} to
analytical queries over cold data~\cite{lambada,starling}.
These data management workloads benefit from serverless computing by taking advantage of the
unlimited elasticity, pay per use,
and lower start-up and set-up overhead
provided by a serverless infrastructure.


\vspace{-0.5em}
\paragraph*{\underline{\bf Serverless Computing and FaaS}}
Serverless computing has been offered by major cloud service providers (e.g., AWS Lambda~\cite{lambda}, Azure Functions~\cite{azure}, Google Cloud Functions~\cite{google_function}) and is favored by many applications (e.g., event processing, API composition, API aggregation, data flow control, etc.~\cite{serverless-trends}) as it lifts the burden of provisioning and managing cloud computation resources (e.g., with auto-scaling) from application developers.
Serverless computing also offers a novel ``pay by usage'' pricing model and can be more cost-effective compared to traditional ``serverful'' cloud computing that charges users based on the amount of computation resources being reserved.
With serverless, the user specifies a 
\textit{function} that she hopes to execute and is only charged
for the duration of the function execution. The users
can also easily scale up the computation by
specifying the number of such functions that are executed
concurrently. In this paper, we use the term FaaS (function as a service) to denote the serverless infrastructure
and use the term IaaS (infrastructure as a service)
to denote the VM-based infrastructure.

\vspace{-0.5em}
\paragraph*{\underline{\bf ML and Data Management}}
Modern data management systems 
are increasingly tightly integrated with advanced
analytics such as data mining and machine learning (ML).
Today, many database systems support
a variety of machine learning training and inference tasks~\cite{MADlib,SAPHANA,ORE,DB4ML,TeraSQLML}. 
Offering ML functionalities inside a database system reduces
data movement across system boundaries and makes it
possible for the ML components to take advantage of
built-in database mechanisms such as access control and integrity 
checking. Two important aspects 
of integrating machine learning into DBMS are \textit{performance} and \textit{scalability}. As a result, the database 
community has been one of the driving forces 
behind recent advancement of distributed machine
learning~\cite{MADlib,RDBMS_analytics,HybridParallelVLDB,FlexpsVLDB,CompressLinear,VerticaML,MLlib*,Keystoneml,HeteroSIGMOD,DBS-062,tictac,comm_schedule}.

\vspace{-0.5em}
\paragraph*{\underline{\bf Motivation: FaaS Meets ML Training}}
Inspired by these two emerging technological trends, in this paper we focus on one of their intersections by enabling distributed ML \emph{training} on top of serverless computing.
While FaaS is a natural choice for ML inference~\cite{ServeServerless}, it is unclear whether FaaS can also be beneficial when it comes to ML training.
Indeed, this is a nontrivial problem and there has been active recent research from both the academia and the industry.
For example, AWS provides one example of serverless ML training in AWS Lambda using SageMaker and AutoGluon~\cite{aws-serverless-ml}.
Such supports are useful when building 
``training-as-a-service platforms''
in which requests of ML model training
come in continuously from multiple users or
one model is continuously re-trained when new training data arrives,
and are also useful when providing users with ``code-free'' experience of ML 
training without worrying about managing the underlying infrastructure. 
Not surprisingly,
training ML models using serverless 
infrastructure has also attracted increasingly
intensive attention from the academia~\cite{serverless-CIDR19,cirrus,Oversketched,SIREN,carreira2018serverless,ServerlessNN,Stratum}.
We expect to see even more applications and 
researches focusing on training ML models using FaaS
infrastructures in the near future.

Our goal in this work is to
\textit{understand the system tradeoff of supporting distributed ML training with serverless 
infrastructures.}
Specifically, we are interested in the following question:
\begin{quote}
\em
When can a serverless infrastructure
(FaaS) outperform a VM-based, ``serverful'' infrastructure (IaaS) for distributed ML training?
\end{quote}

\vspace{-0.5em}
\paragraph*{\underline{\bf State of the Landscape}}
Despite of 
these recent interests, these 
early explorations  
depict a ``mysterious'' picture 
of the the 
relative performance of 
IaaS and FaaS for training
ML models. Although previous 
work~\cite{serverless-CIDR19,cirrus,Oversketched,SIREN} has illustrated 
up to two orders of magnitude 
performance improvements 
of FaaS over IaaS
over a diverse range of workloads,
the conclusion remains inconclusive.
In many of these early explorations, 
FaaS and IaaS are often not put onto the
same ground for comparison
(see Section~\ref{sec:related-work}): either 
the IaaS or FaaS implementations could
be further optimized or 
only micro-benchmarking is conducted. Moreover,
similar to other non-ML workloads,
we expect a delicate system 
tradeoff in which FaaS only outperforms IaaS
in specific regimes~\cite{lambada,ShuffleServerless,starling}. However, a systematic depiction of this tradeoff space, with an analytical 
model, is largely lacking for ML training.

\vspace{-0.5em}
\paragraph*{\bf \underline{Summary of Contributions}} 
In this paper, we conduct an extensive experimental study inspired by the current landscape of FaaS-based distributed ML training.
Specifically, we
\begin{quote}
\em systematically explore both the
\underline{algorithm choice} and 
\underline{system design} for
both FaaS
and IaaS ML training strategies and 
depict the tradeoff
over a diverse range of ML models,
training workloads, and infrastructure choices.
\end{quote}
In addition to the depiction of this empirical 
tradeoff using today's infrastructure, we further
\begin{quote}
\em develop an \underline{analytical model} that characterizes
the tradeoff between FaaS and IaaS-based 
training, and use it to speculate performances of potential configurations used by \emph{future} systems.
\end{quote}

In designing our experimental study, we 
follow a set of principles for a fair comparison 
between FaaS and IaaS:

\begin{enumerate}
\item 
{\bf Fair Comparison.} 
To provide an objective benchmark and evaluation of Faas and IaaS, we stick to 
the following principled methodology 
in this empirical study:
(1) both FaaS and IaaS implement
the \textit{same} set of algorithms 
(e.g., SGD and ADMM) to avoid 
apple-to-orange comparisons 
such as comparing FaaS running ADMM with
IaaS running SGD; and (2) we compare
FaaS and IaaS running the most suitable algorithms
with the most suitable 
hyper-parameters such as VM type and
number of workers.
%
\item {\bf End-to-end Benchmark.}
We focus on the end-to-end
training performance -- the wall clock time (or cost in dollar)
that each system needs to converge to the \textit{same} loss.
%
\item {\bf Strong IaaS Competitors and 
Well-optimized FaaS System.}
We use state-of-the-art  
systems as the IaaS solution, which 
are often much faster than what has
been used in previous work
showing FaaS is faster than IaaS. We also conduct careful system 
optimizations and designs for FaaS. The prototype 
system, LambdaML, can often pick 
a point in the design space that is faster than what has
been chosen by previous FaaS systems.
%
\end{enumerate}

\vspace{-0.5em}
\paragraph*{\bf \underline{Summary of Insights}} Our study leads to two 
key insights:

\begin{enumerate}
\item {\textit{FaaS can be faster than IaaS, but only in a specific regime}: when the 
underlying workload can be made communication efficient, in terms of both convergence and amount of data communicated.
On one hand, there \textit{exists} realistic datasets and models
that can take advantage of this benefit;
on the other hand, there are also workloads 
under which FaaS performs significantly worse. 
}
\item {\textit{When FaaS is much faster, it is not much cheaper}: One insight that holds 
across all scenarios is that even when FaaS
is much faster than IaaS, it usually incurs
a comparable cost in dollar. This 
mirrors the results for other workloads
in Lambada~\cite{lambada}
and Starling~\cite{starling}, illustrating
the impact of FaaS pricing model.
}
\end{enumerate}

\vspace{0.5em}
\noindent
{\bf(Paper Organization)}
We start by an overview of existing distributed ML technologies and FaaS offerings (Section~\ref{sec:preliminaries}).
We then turn to an anatomy of the design space of FaaS-based ML systems, following which we implement a Serverless ML platform called \system (Section~\ref{sec:serverless}).
We present an in-depth evaluation 
of various design options 
when implementing \system (Section~\ref{sec:eval:lambdaml}).
We further present a systematic study of FaaS-based versus IaaS-based ML systems, both empirically and analytically (Section~\ref{sec:serverless-versus-serverful}). 
We summarize related work in Section~\ref{sec:related-work} and conclude in Section~\ref{sec:conclusion}.

\vspace{0.5em}
\noindent
{\bf(Reproducibility and Open Source Artifact)}
LambdaML is publicly available at \url{https://github.com/DS3Lab/LambdaML}. All experiments can be reproduced following the instructions at \url{https://github.com/DS3Lab/LambdaML/blob/master/reproducibility.md}.

\section{Preliminaries}\label{sec:preliminaries}

In this section, we present a brief overview of state-of-the-art distributed ML technologies, as well as the current offerings of FaaS (serverless) infrastructures by major cloud service providers. 

\vspace{-0.5em}
\subsection{Distributed Machine Learning}

\subsubsection{Data and Model}

A training dataset $D$ consists of $n$ i.i.d. data examples that are generated by the underlying data distribution $\mathcal{D}$.
Let $D=\{(\mathbf{x}_i\in \mathbb{R}^{n}, y_i\in\mathbb{R})\}_{i=1}^N$, where $\mathbf{x}_i$ represents the \emph{feature vector} and $y_i$ represents the \emph{label} of the $i^{\text{th}}$ data example.
The goal of ML training is to find an ML model $\mathbf{w}$ that minimizes a \emph{loss function} $f$ over the training dataset $D$:
$\arg\min_{\mathbf{w}} \frac{1}{N} \sum\nolimits_i f(\mathbf{x}_i, y_i, \mathbf{w})$.

\subsubsection{Optimization Algorithm}

Different ML models rely on different optimization algorithms.
Most of these optimization algorithms are \emph{iterative}.
In each iteration, the training procedure would typically scan the training data, compute necessary quantities (e.g., gradients), and update the model.
This iterative procedure terminates/converges when there are no more updates to the model.
During the training procedure, each pass over the entire data is called an \emph{epoch}.
For instance, mini-batch stochastic gradient descent (SGD) processes one batch of data during each iteration, and thus one epoch involves multiple iterations; on the other hand, k-means processes all data, and thus one epoch, in each iteration.

\noindent\textbf{(Distributed Optimization)} When a single machine does not have the computation power or storage capacity (e.g., memory) to host an ML training job, one has to deploy and execute the job across multiple machines.
Training ML models in a distributed setting is more complex, due to the extra complexity of distributed computation as well as coordination of the communication between executors.
Lots of distributed optimization algorithms have been proposed.
Some of them are straightforward extensions of their single-node counterparts (e.g., k-means), while the others require more sophisticated adaptations dedicated to distributed execution environments (e.g., parallelized SGD~\cite{ZinkevichWSL10-psgd}, distributed ADMM~\cite{ADMM}).

\vspace{-0.5em}
\subsubsection{Communication Mechanism}

One key differentiator in the design and implementation of distributed ML systems is the communication mechanism employed.
In the following, we present a brief summary of communication mechanisms leveraged by existing systems, with respect to a simple taxonomy regarding \emph{communication channel}, \emph{communication pattern}, and \emph{synchronization protocol}.

\noindent\textbf{(Communication Channel)}  The efficiency of data transmission relies on the underlying communication channel. 
While one can rely on pure \emph{message passing} between executors, this shared-nothing mechanism may be inefficient in many circumstances.
For example, when running SGD in parallel, each executor may have to broadcast its local versions of global states (e.g., gradients, model parameters) to every other executor whenever a synchronization point is reached.
As an alternative, one can use certain storage medium, such as a disk-based file system or an in-memory key-value store, to provide a central access point for these shareable global states. 

\noindent\textbf{(Communication Pattern)} 
A number of collective communication primitives can be used for data exchange between executors~\cite{kungfu_osdi}, such as \texttt{Gather}, \texttt{AllReduce}, and \texttt{ScatterReduce}.

\noindent\textbf{(Synchronization Protocol)} The iterative nature of the optimization algorithms may imply certain dependencies across successive iterations, which force synchronizations between executors at certain boundary points~\cite{ada_comm}.
A synchronization protocol has to be specified regarding when such synchronizations are necessary.
Two common protocols used by existing systems are \emph{bulk synchronous parallel} (BSP) and \emph{asynchronous parallel} (ASP).
BSP is preferred if one requires certain convergence
or reproducibility guarantees, where no work can proceed to the next iteration without having \emph{all} workers finish the current iteration.
In contrast, ASP does not enforce such synchronization barriers, but 
could potentially hurt the empirical convergence
rate in some applications.

\vspace{-0.5em}
\subsection{FaaS vs. IaaS for ML}

Most of the aforementioned distributed ML technologies have only been applied in \emph{IaaS} environments on cloud, where users have to build a cluster by renting VMs or reserve a cluster with predetermined configuration parameters (e.g., Azure HDInsight~\cite{HDInsight}).
As a result, users pay bills based on the computation resources that have been reserved, regardless of whether these resources are in use or not.
Moreover, users have to manage the resources by themselves---there is no elasticity or auto-scaling if the reserved computation resources turn out to be insufficient, even for just a short moment (e.g., during the peak of a periodic or seasonal workload).
Therefore, to tolerate such uncertainties, users tend to \emph{overprovisioning} by reserving more computation resources than actually needed.

The move towards FaaS infrastructures lifts the burden of managing computation resources from users.
Resource allocation in FaaS is on-demand and auto-scaled, and users are only charged by their actual resource usages.
The downside is that FaaS currently does not support customized scaling and scheduling strategies.
Although the merits of FaaS are very attractive, 
current offerings by major cloud service providers (e.g., AWS Lambda~\cite{lambda}, Azure Functions~\cite{azure}, Google Cloud Functions~\cite{google_function}) impose certain limitations and/or constraints that shed some of the values by shifting from IaaS to FaaS infrastructures.
Current FaaS infrastructures only support \emph{stateless} function calls with limited computation resource and duration.
For instance, a function call in AWS Lambda can use up to 3GB of memory and must finish within 15 minutes~\cite{aws-lambda-limits}.
Such constraints automatically eliminate some simple yet natural ideas on implementing FaaS-based ML systems.
For example, one cannot just wrap the code of SGD in an AWS Lambda function and execute it, which would easily run out of memory or hit the timeout limit on large training data.
Indeed, state-of-the-art FaaS systems raise lots of new challenges for designing ML systems and leads to a rich design space, as we shall cover in the next section.

\begin{figure}[t]
  \centering
  \includegraphics[width=0.75\columnwidth]{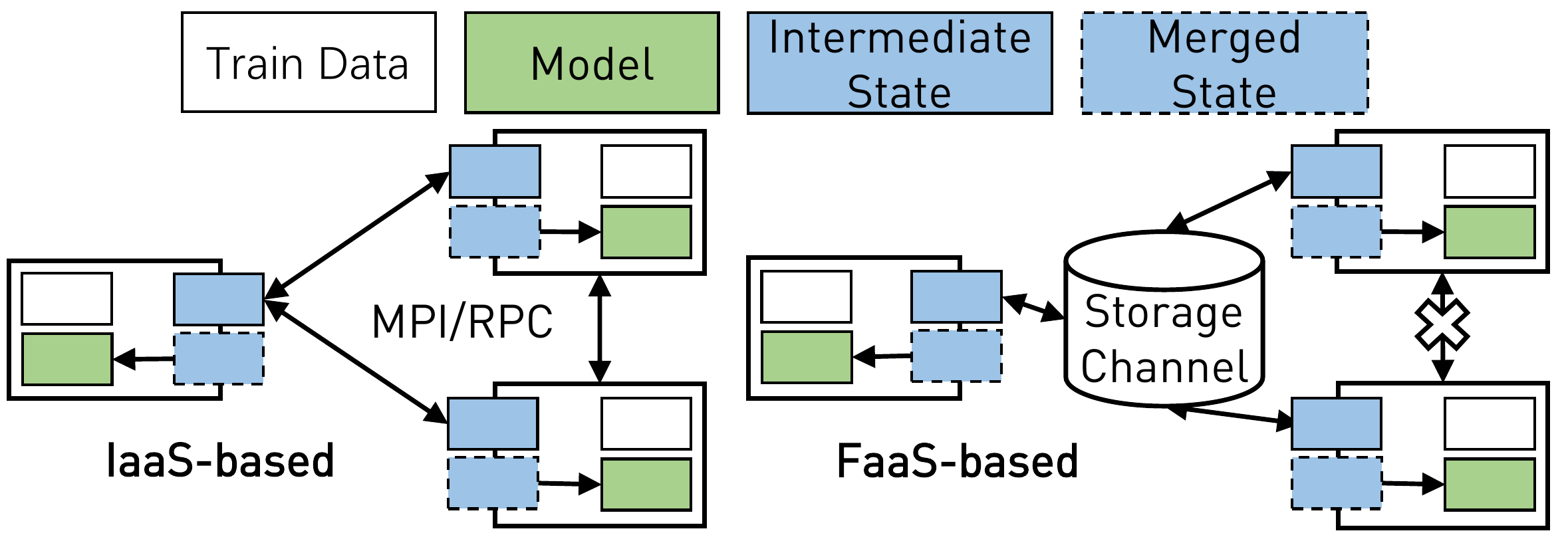}
  \vspace{-1em}
  \caption{IaaS vs. FaaS-based ML system architectures.}
  \label{fig:comm_vs_serverless}
\end{figure}

\vspace{-0.5em}
\section{LambdaML}\label{sec:serverless}


We implement \system, a prototype FaaS-based ML system built on top of Amazon Lambda, and
study the trade-offs in training ML models over serverless infrastructure.


\vspace{-0.5em}
\subsection{System Overview}
\label{sec:serverless:design}

\begin{figure}[t]
  \centering
  \includegraphics[width=0.9\columnwidth]{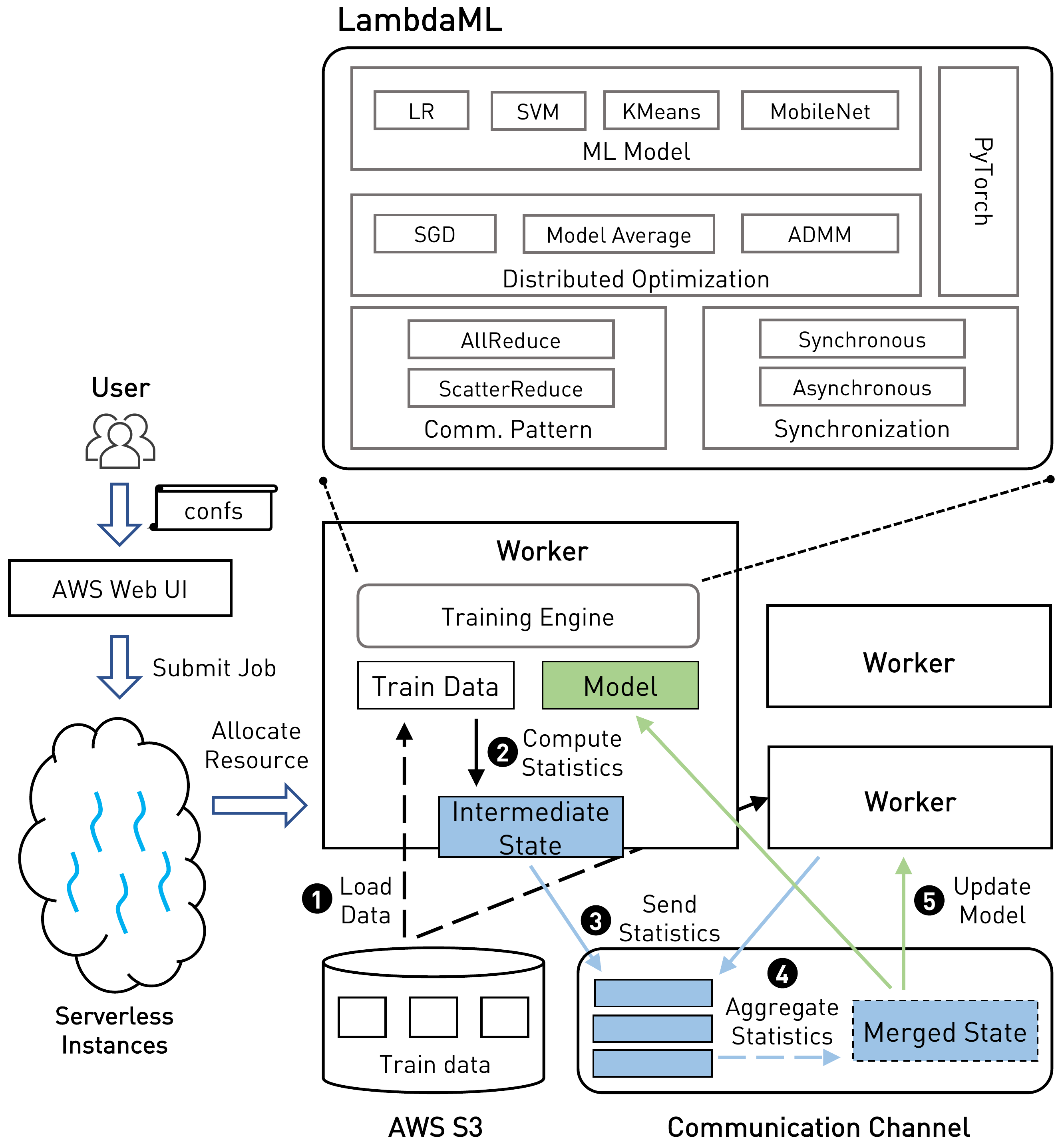}
  \vspace{-1em}
  \caption{Framework of \system.}
  \label{fig:lambdaml_framework}
\end{figure}

\noindent\textbf{(Challenges)}
As mentioned in Section~\ref{sec:preliminaries}, we need to consider four dimensions when developing distributed ML systems: (1) the \emph{distributed optimization algorithm}, (2) the \emph{communication channel}, (3) the \emph{communication pattern}, and (4) the \emph{synchronization protocol}.
These elements remain valid when migrating ML systems from IaaS to FaaS infrastructures, though new challenges arise.
One main challenge is that current FaaS infrastructures do not allow direct communication between stateless functions.
As a result, one has to use certain storage channel to allow the functions to read/write intermediate state information generated during the iterative training procedure.
Figure~\ref{fig:comm_vs_serverless} highlights this key difference between IaaS and FaaS designs of ML training systems.

\vspace{0.3em}
\noindent\textbf{(Framework of \system)}
Figure~\ref{fig:lambdaml_framework} shows the framework of \system.
When one user specifies the training configurations in AWS web UI (e.g., data location, resources, optimization algorithm, and hyperparameters), AWS \underline{\em submits job} to the serverless infrastructure that \underline{\em allocates resources} (i.e., serverless instances) according to the user request.
Each running instance is a \underline{\em worker} in \system.
The \underline{\em training data} is partitioned and stored in S3, a distributed storage service in AWS.
Each worker maintains a \underline{\em local model copy} and uses the library of \system to train a machine learning model.

\vspace{0.3em}
\noindent\textbf{(Job Execution)}
A training job in \system has the steps below:

\begin{enumerate}
    \item{\em Load data.}
    Each worker loads the corresponding partition of training data from S3.
    
    \item{\em Compute statistics.}
    Each worker creates the ML model with PyTorch and computes statistics for aggregation using the training data and the local model parameters.
    Different optimization algorithms may choose different statistics for aggregation, e.g., gradient in distributed SGD and local model in distributed model averaging
    (see Section~\ref{sec:design_opt_algorithm} for more details).
    
    \item{\em Send statistics.}
    In a distributed setting, the statistics are sent to a communication channel
    (see Section~\ref{sec:design_communication_channel}).
    
    \item{\em Aggregate statistics.}
    The statistics from all the workers, which are considered as intermediate states, are aggregated using a certain pattern, generating a global state of the statistics (see Section~\ref{sec:design_comm_pattern}).
    
    \item{\em Update model.}
    Each worker gets the merged state of the statistics, with which the local model is updated.
    For an iterative optimization algorithm, 
    if one worker is allowed to proceed according to a synchronization protocol (see Section~\ref{sec:design_sync_protocol}), it goes back to step (2) and runs the next iteration.
    
\end{enumerate}

\vspace{0.3em}
\noindent
\textbf{(Components of \system Library)}
As shown in Figure~\ref{fig:lambdaml_framework},
the computation library of \system is developed on top of PyTorch, which provides a wide range of ML models and \texttt{autograd} functionalities.
To run \underline{\em distributed optimization algorithms}, the communication library of \system relies on some \underline{\em communication channel} to aggregate local statistics via certain
\underline{\em communication pattern} and governs the iterative process using a \underline{\em synchronization protocol.}

\vspace{-0.5em}
\subsection{Implementation of \system}

In this section we elaborate the aforementioned four major aspects in the implementation of \system ---
distributed optimization algorithm, communication channel, communication pattern, and synchronization protocol.
Each aspect contains a rich design space which should be studied carefully.

\vspace{-0.5em}
\subsubsection{Distributed Optimization Algorithm}
\label{sec:design_opt_algorithm}

In our study, we focus on the following distributed optimization algorithms.

\vspace{0.3em}
\noindent\textbf{(Distributed SGD)}
Stochastic gradient descent (SGD) is perhaps the most popular optimization algorithm in today's world, partly attributed to the success of deep neural networks.
We consider two variants when implementing SGD in a distributed manner: (1) {\em gradient averaging} (GA) and (2) {\em model averaging} (MA).
In both implementations, we partition training data evenly and have one executor be in charge of one partition.
Each executor runs mini-batch SGD independently and in parallel, while sharing and updating the global ML model at certain synchronization barriers (e.g., after one or a couple of iterations).
The difference lies in the way that the global model gets updated.
GA updates the global model in \emph{every} iteration by harvesting and aggregating the (updated) gradients from the executors.
In contrast, MA collects and aggregates the (updated) local models, instead of the gradients, from the executors and do not force synchronization at the end of each iteration.
That is, executors may combine the local model updates accumulated in a number of iterations before synchronizing with others to obtain the latest consistent view of the global model.
We refer readers to~\cite{MLlib*} for a more detailed comparison between GA and MA.

\vspace{0.3em}
\noindent\textbf{(Distributed ADMM)}
Alternating direction method of multipliers (ADMM) is another popular distributed optimization algorithm proposed by Boyd et al.~\cite{ADMM}.
ADMM breaks a large-scale convex optimization problem into several smaller subproblems, each of which is easier to handle.
In the distributed implementation of \system, each executor solves one subproblem (i.e., until convergence of the local solution) and then exchanges local models with other executors to obtain the latest view of the global model.
While this paradigm has a similar pattern as model averaging, it has been shown that ADMM can have better convergence guarantees~\cite{ADMM}.

\begin{figure}[t]
  \centering
  \includegraphics[width=0.75\columnwidth]{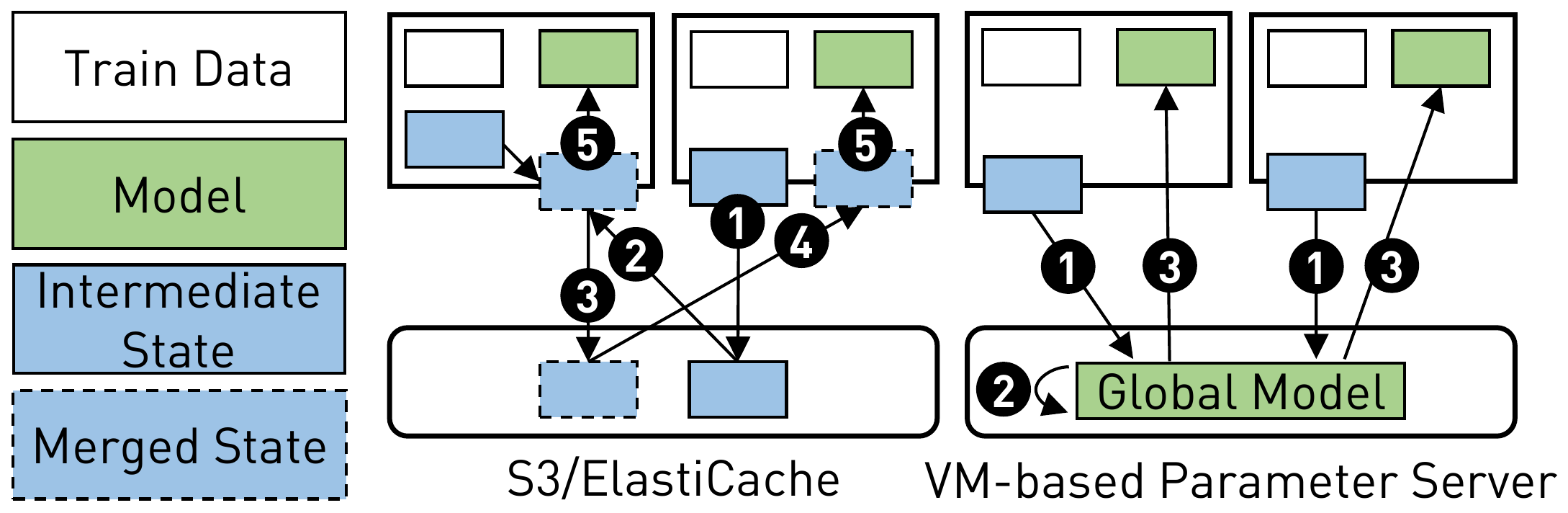}
  \vspace{-1em}
  \caption{An FaaS-based data aggregation.}
  \label{fig:implement_medium}
\end{figure}

\vspace{-0.5em}
\subsubsection{Communication Channel} \label{sec:design_communication_channel}
As we mentioned, it is necessary to have a storage component in an FaaS-based ML system to allow stateless functions to read/write intermediate state information generated during the lifecycle of ML training.
With this storage component, we are able to aggregate data between running instances in the implementation of distributed optimization algorithms.
Often, there are various options for this storage component, with a broad spectrum of cost/performance tradeoffs.
For example, in Amazon AWS, one can choose between four alternatives---S3, ElastiCache for Redis, ElastiCache for Memcached, and DynamoDB.
S3 is a disk-based object storage service~\cite{s3}, whereas Redis and Memcached are in-memory key-value data stores provided by Amazon ElastiCache~\cite{elasticache}.
DynamoDB is an in-memory key-value database hosted by Amazon AWS~\cite{aws_dynamodb}.
In addition to using external cloud-based storage services, one may also consider building his/her own customized storage layer.
For instance, Cirrus~\cite{cirrus} implements a parameter server~\cite{HeteroSIGMOD} on top of a virtual machine (VM), which serves as the storage access point of the global model shared by the executors (implemented using AWS Lambda functions).
This design, however, is not a pure FaaS architecture, as one has to maintain the parameter server by himself.
We will refer to it as a \emph{hybrid} design.

Different choices on communication channel lead to different cost/performance tradeoffs.
For example, on AWS, it usually takes some time to start an ElastiCache instance or a VM, whereas S3 does not incur such a startup delay since it is an ``always on'' service.
On the other hand, accessing files stored in S3 is in general slower but cheaper than accessing data stored in ElastiCache.

\vspace{0.3em}
\noindent{\bf (An FaaS-based Data Aggregation)}
We now design a communication scheme for data aggregation using a storage service, such as S3 or ElastiCache, as the communication channel.
As shown in Figure~\ref{fig:implement_medium}, the entire communication process contains the following steps: 1) Each executor stores its generated intermediate data as a temporary file in S3 or ElastiCache; 
2) The first executor (i.e., the \emph{leader}) pulls all temporary files from the storage service and merges them to a single file; 3) The leader writes the merged file back to the storage service; 4) All the other executors (except the leader) read the merged file from the storage service; 5) All executors refresh their (local) model with information read from the merged file.

Figure~\ref{fig:implement_medium} also presents an alternative implementation using a VM-based parameter server as in the hybrid design exemplified by Cirrus~\cite{cirrus}.
In this implementation, 1) each executor pushes local updates to the parameter server, with which 2) the parameter server further updates the global model.
Afterwards, 3) each executor pulls the latest model from the parameter server.

\vspace{-0.5em}
\subsubsection{Communication Pattern}
\label{sec:design_comm_pattern}

\begin{figure}[t]
  \centering
  \includegraphics[width=0.65\columnwidth]{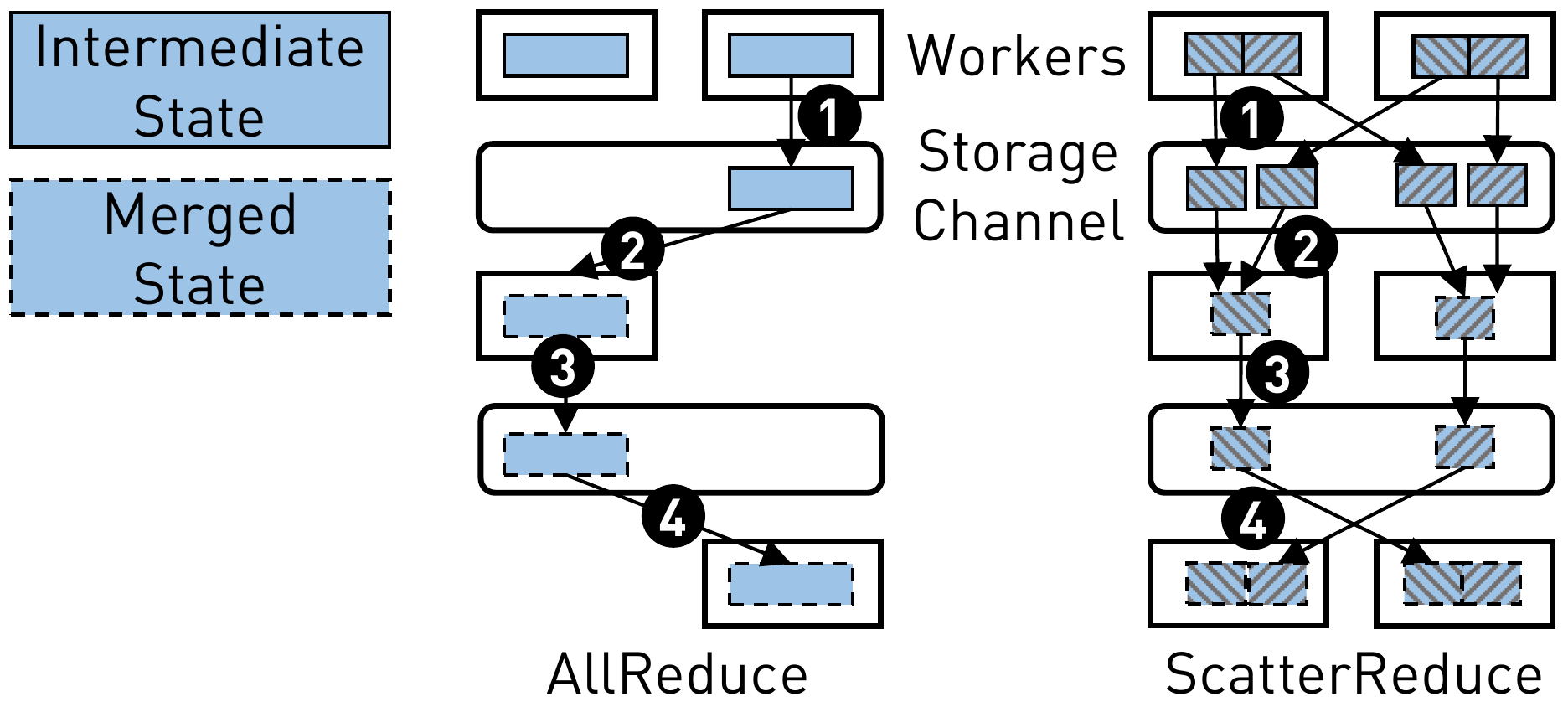}
  \vspace{-1em}
  \caption{AllReduce vs. ScatterReduce.}
  \label{fig:implement_pattern}
\end{figure}

To study the impact of communication patterns, we focus on two MPI primitives, \texttt{AllReduce} and \texttt{ScatterReduce}, that have been widely implemented in distributed ML systems~\cite{MLlib*}.
Figure~\ref{fig:implement_pattern} presents the high-level designs of \texttt{AllReduce} and \texttt{ScatterReduce} in an FaaS environment with an external storage service such as S3 or ElastiCache. 

\vspace{0.3em}
\noindent\textbf{(AllReduce)}
With \texttt{AllReduce}, all executors first write their local updates to the storage.
Then the first executor (i.e., the leader) reduces/aggregates the local updates and writes the aggregated updates back to the storage service. 
Finally, all the other executors read the aggregated updates back from the storage service.

\vspace{0.3em}
\noindent\textbf{(ScatterReduce)}
When there are too many executors or a large amount of local updates to be aggregated, the single leader executor in \texttt{AllReduce} may become a performance bottleneck.
This is alleviated by using \texttt{ScatterReduce}.
Here, all executors are involved in the reduce/aggregate phase, each taking care of one partition of the local updates being aggregated.
Specifically, assume that we have $n$ executors.
Each executor divides its local updates into $n$ partitions, and then writes each partition separately (e.g., as a file) to the storage service.
During the reduce/aggregate phase, the executor $i$ ($1\leq i\leq n$) collects the $i^\text{th}$ partitions generated by all executors and aggregates them.
It then writes the aggregated result back to the storage service.
Finally, each executor $i$ pulls aggregated results produced by all other executors to obtain the entire model.

\begin{figure}[!t]
  \centering
  \includegraphics[width=0.8\columnwidth]{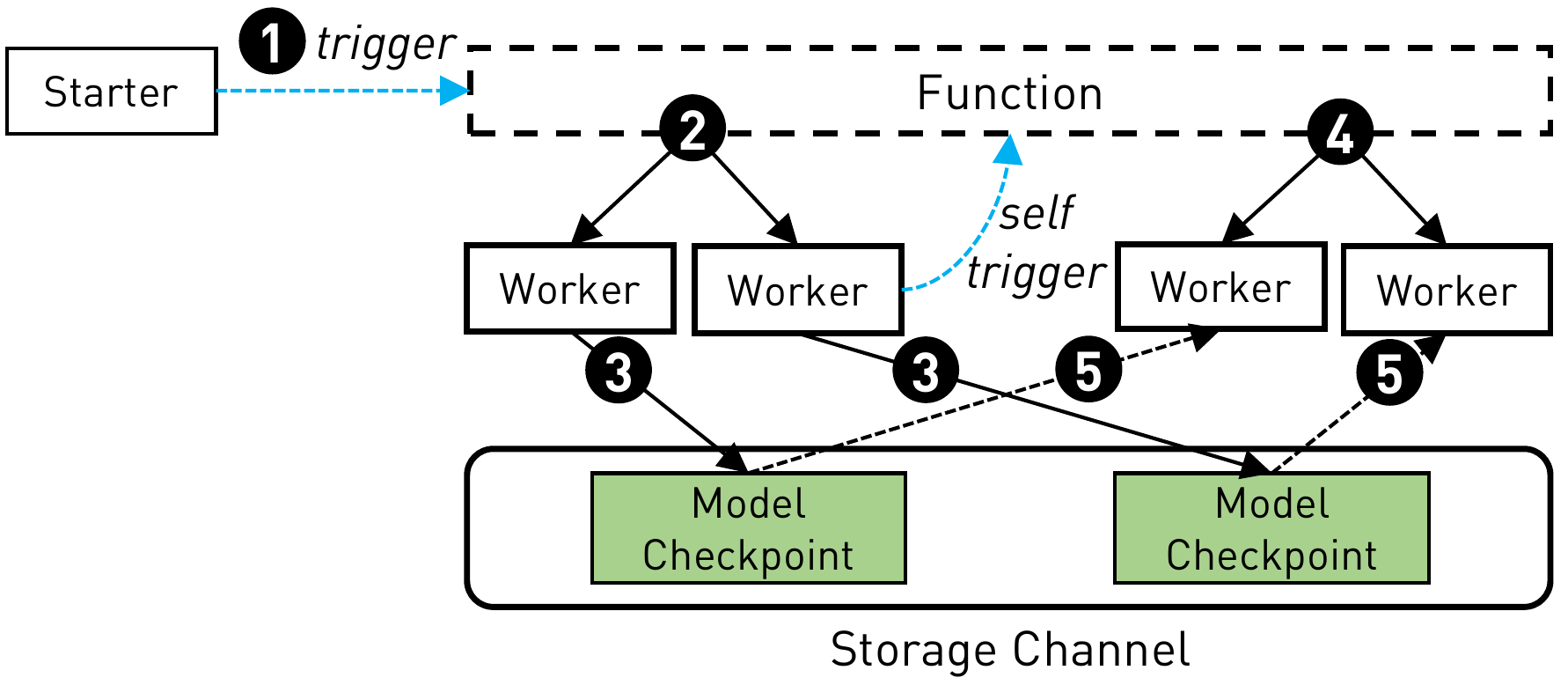}
  \vspace{-1em}
  \caption{Invocation structure of Lambda workers.}
  \label{fig:invocation}
\end{figure}

\vspace{-0.5em}
\subsubsection{Synchronization Protocol}
\label{sec:design_sync_protocol}

We focus on two synchronization protocols that have been adopted by many existing distributed ML systems.
One can simply implement these protocols on top of serverful architectures by using standard primitives of message passing interface (MPI), such as \textit{MPI\_Barrier}.
Implementations on top of FaaS architectures, however, are not trivial, given that stateless functions cannot directly communicate with each other.

\vspace{0.2em}
\noindent\textbf{(Synchronous)}
We design a two-phase synchronous protocol, which includes a merging and an 
updating phase. We illustrate this in FaaS architecture that leverages an external storage service:

\begin{itemize}
    \item{\em Merging phase.} All executors first write their local updates to the storage service.
    The reducer/aggregator (e.g., the leader in \texttt{AllReduce} and essentially every executor in \texttt{ScatterReduce}) then needs to make sure that it has aggregated local updates from all other executors. Otherwise it should just wait.
    \item{\em Updating phase.} After the aggregator finishes aggregating all data and stores the aggregated information back to the storage service, all executors can read the aggregated information to update their local models and then proceed with next round of training.
\end{itemize}
All executors are synchronized using this two-phase framework.
Moreover, one can rely on certain atomicity guarantees provided by the storage service to implement these two phases.
Here we present the implementation of our proposed synchronous protocol.

\begin{itemize}
    \item {\em Implementation of the Merging Phase.}
We name the files that store local model updates using a scheme that includes all essential information, such as the training epoch, the training iteration, and the partition ID.
The reducer/aggregator can then request the list of file names from the storage service (using APIs that are presumed \emph{atomic}), filter out uninteresting ones, and then count the number of files that it has aggregated. When the number of aggregated files equals the number of workers, the aggregator can proceed. Otherwise, it should wait and keep polling the storage service until the desired number of files is reported.

    \item{\em Implementation of the Updating Phase.}
We name the merged file that contains the aggregated model updates in a similar manner, which consists of the training epoch, the training iteration, and the partition ID.
For an executor that is pending on the merged file, it can then keep polling the storage service until the name of the merged file shows up.
\end{itemize}

\vspace{0.2em}
\noindent\textbf{(Asynchronous)}
Following the approach of SIREN~\cite{SIREN}, the implementation of asynchronous communication is simpler.
One replica of the trained model is stored on the storage service as a global state.
Each executor runs independently -- it reads the model from the storage service, updates the model with training data, writes the new model back to the storage service -- without caring about the speeds of the other executors.

\vspace{-0.5em}
\subsection{Other Implementation Details}
\label{sec:serverless:impl}

This section provides the additional implementation details of \system that are relevant for understanding its performance.

\vspace{-0.5em}
\subsubsection{Handling Limited Lifetime}
One major limitation of Lambda functions is their (short) lifetime,
that is, the execution time cannot be longer than 15 minutes.
We implement a \emph{hierarchical invocation} mechanism to schedule their executions, as illustrated in Figure~\ref{fig:invocation}.
Assume that the training data has been \emph{partitioned} and we have one executor (i.e., a Lambda function) for each partition.
We start Lambda executors with the following steps: (1) a {\em starter} Lambda function is triggered once the training data has been uploaded into the storage service, e.g., S3; (2) the starter triggers $n$ {\em worker} Lambda functions where $n$ is the number of partitions of the training data.
Each worker is in charge of its partition, which is associated with metadata such as the path to the partition file and the ID of the partition/worker.
Moreover, a worker monitors its execution to watch for the 15-minute timeout.
It pauses execution when the timeout is approaching, and saves a checkpoint to the storage service that includes the latest local model parameters.
It then resumes execution by triggering its Lambda function with a new worker.
The new worker inherits the same worker ID and thus would take care of the same training data partition (using model parameters saved in the checkpoint).

\noindent{\bf (Limitation)}
Under the current design, this mechanism cannot 
support the scenario in which 
\emph{a single iteration} takes longer than 15 minutes.
We have not observed such a workload in our evaluation,
and it would require a very large
model and batch size for a single iteration to exceed 15 minutes.
Especially given the memory 
limitation (3GB) of FaaS, we do not expect this to happen in most realistic scenarios.
A potential workaround to accommodate this limitation is to use a smaller batch size so that one iteration takes less than 15 minutes.
A more fundamental solution might be to
explore model-parallelism~\cite{singa,alexnet} in the context of FaaS, which is an interesting direction for future research.

\section{Evaluation of LambdaML}
\label{sec:eval:lambdaml}

We evaluate \system with the goal of comparing the various design options that have been covered in Section~\ref{sec:serverless}.
We start by laying out the experiment settings and then report evaluation results with respect to each dimension of the design space.

\vspace{-0.5em}
\subsection{Experiment Settings}
\label{sec:eval_lambda_setting}

\vspace{0.2em}
\noindent{\bf (Datasets)}
Figure~\ref{fig:micro_dataset} presents the datasets used in our evaluation.
In this section, we focus on smaller datasets to understand the system
behavior and leave
the larger datasets (YFCC100M and Criteo)
to the next section when we conduct the end-to-end study.
We focus on {\bf Higgs}, {\bf RCV1} and {\bf Cifar10}.
{\bf Higgs} is a dataset for binary classification, produced by using Monte Carlo simulations.
{\bf Higgs} contains 11 million instances, and each instance has 28 features.
{\bf RCV1} is a two-class classification corpus of manually categorized newswire stories made available by Reuters~\cite{rcv1}.
The feature of each training instance is a 47236-dimensional sparse vector, in which every value is a normalized TF-IDF value.
{\bf Cifar10} is an image dataset that consists of 60 thousand 32$\times$32 images categorized by 10 classes, with 6 thousand images per class.

\begin{figure}[!t]
  \centering
  \subfloat[\small Micro benchmark.]{
    \scriptsize
    \label{fig:micro_dataset}
    \begin{tabular}{c r r r }
    \hline
    Dataset & Size & \# Ins & \# Feat \\
    \hline
    Cifar10 & 220 MB & 60 K & 1 K  \\
    RCV1 & 1.2 GB & 697 K & 47 K   \\
    Higgs & 8 GB & 11 M & 28 \\
    \hline
    \end{tabular}}
  \subfloat[\small End-to-end benchmark.]{
    \scriptsize
    \label{fig:e2e_dataset}
    \begin{tabular}{c r r r }
    \hline
    Dataset & Size & \# Ins & \# Feat \\
    \hline
    Cifar10 & 220 MB & 60 K & 1 K  \\
    YFCC100M & 110 GB & 100 M & 4 K \\
    Criteo & 30 GB & 52 M & 1M \\
    \hline
    \end{tabular}}
  \vspace{-1.5em}
  \caption{Datasets used in this work.}
  \label{fig:dataset}
\end{figure}

\vspace{0.2em}
\noindent{\bf (ML Models)}
We use the following ML models in our evaluation.
Logistic Regression ({\bf LR}) and Support Vector Machine ({\bf SVM}) are linear models for classification that are trained by mini-batch SGD or ADMM.
The number of the model parameters is equal to that of input features.
MobileNet ({\bf MN}) is a neural network model that uses depth-wise separable convolutions to build lightweight deep neural networks.
The size of each input image is $224\times 224 \times 3$, and the size of model parameters is 12MB.
ResNet50 ({\bf RN}) is a famous neural network model that was the first to introduce identity mapping and shortcut connection.
KMeans ({\bf KM}) is a clustering model for unsupervised problems, trained by \emph{expectation maximization} (EM).

\vspace{0.3em}
\noindent{\bf (Protocols)}
We randomly shuffle and split the data into a training set (with 90\% of the data) and a validation set (with 10\% of the data).
We report the number 
for {\bf Higgs}
with batch size 100K,
while setting it as 
10K or 1M will not change 
the insights and conclusions;
whereas it is 128 for {\bf MN} and 32 for {\bf RN} over {\bf Cifar10} according to the maximal memory constraint (3GB) of Lambda.
We tune the optimal learning rate for each ML model in the range from 0.001 to 1.
We set a threshold for the observed loss, and we stop training when the threshold is reached.
The threshold is 0.68 for {\bf LR} on {\bf Higgs}, 0.2 for {\bf MN} on {\bf Cifar10}, and 0.1 for {\bf KM} on {\bf Higgs}.

\vspace{0.3em}
\noindent{\bf (Metrics)}
We decouple the system performance into \emph{statistical efficiency} and \emph{system efficiency}.
We measure statistical efficiency by the loss observed over the validation set.
Meanwhile, we measure system efficiency by the execution time of each iteration or epoch.

\vspace{-0.5em}
\subsection{Distributed Optimization Algorithms}

\begin{figure}[t]
  \vspace{-1em}
  \centering
  \subfloat[LR, Higgs.]{
    \label{fig:dist_opt_lr_higgs}
    \includegraphics[width=\columnwidth]{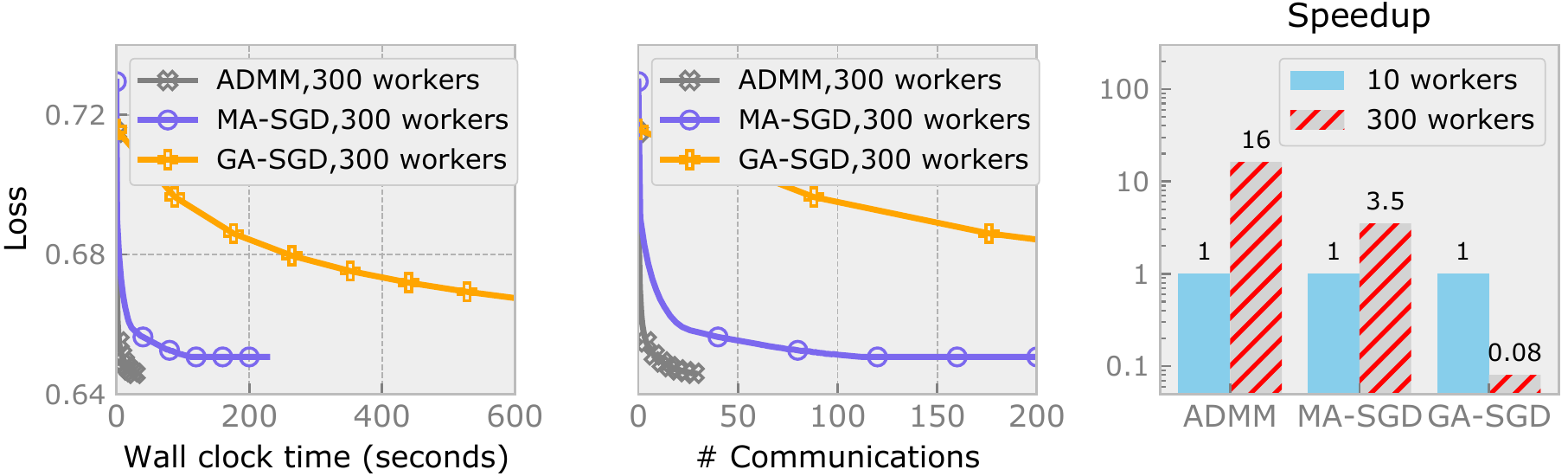}}\\
  \vspace{-1.25em}
  \subfloat[SVM, Higgs.]{
    \label{fig:dist_opt_svm_higgs}
    \includegraphics[width=\columnwidth]{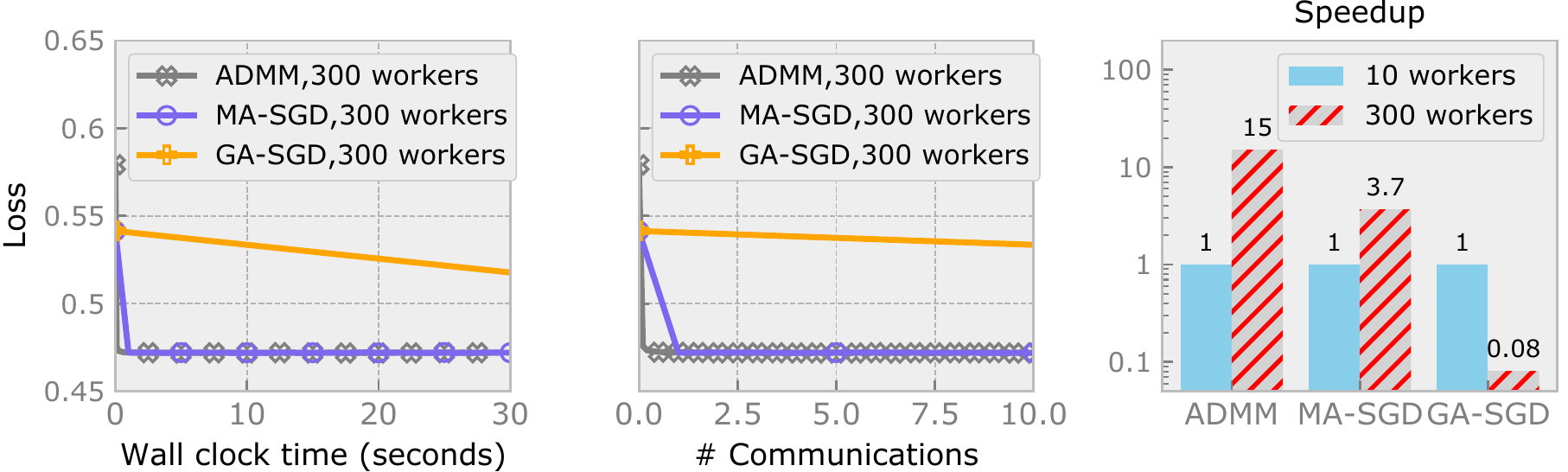}}\\
  \vspace{-1.25em}
  \subfloat[MobileNet, Cifar10.]{
    \label{fig:dist_opt_mn_cifar10}
    \includegraphics[width=\columnwidth]{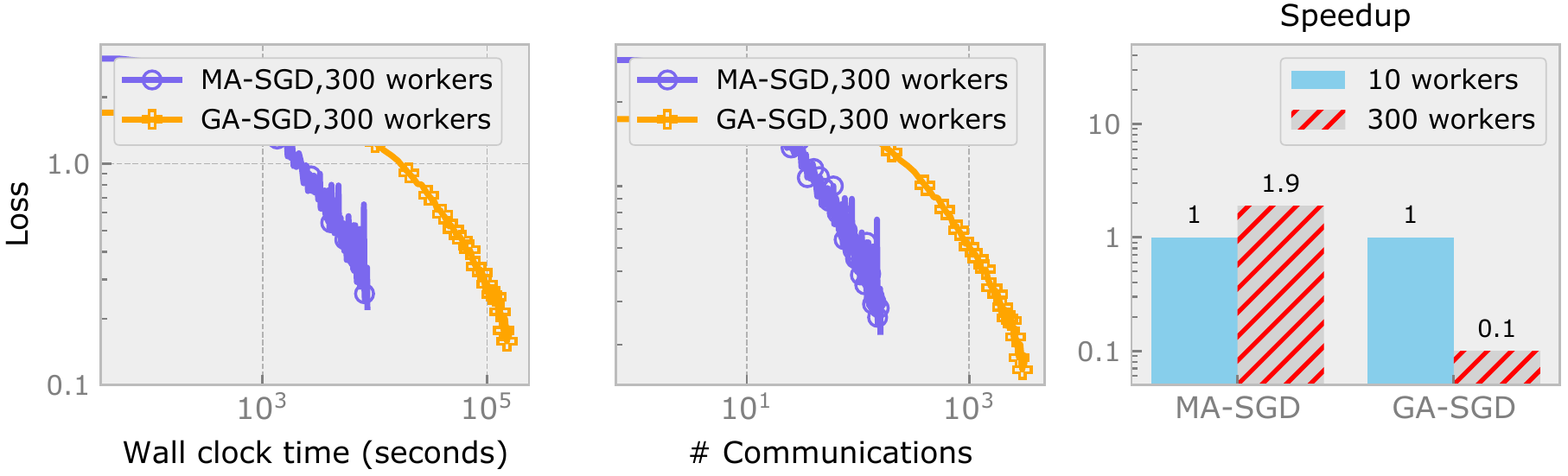}}
  \vspace{-1em}
  \caption{Comparison of distributed optimization methods.}
  \label{fig:dist_opt}
\end{figure}

\begin{quote}
\begin{small}
\em Carefully choosing the right algorithm
goes a long way in optimizing FaaS-based system, and the widely
adopted SGD algorithm is not necessarily ``one-size-fits-all.''
\end{small}
\end{quote}

We implemented GA-SGD (i.e., SGD with gradient averaging), MA-SGD (i.e., SGD with model averaging), and ADMM on top of \system, using ElastiCache for Memcached as the external storage service.
Figure~\ref{fig:dist_opt} presents the results for various data and ML models we tested.
We measure the convergence rate in terms of both the wall clock time and the number of rounds for communication.

\vspace{0.3em}
\noindent{\bf (Results for LR and SVM)}
When training {\bf LR} on {\bf Higgs} using 300 workers, GA-SGD is the slowest because transmitting gradients every batch can lead to high communication cost.
ADMM converges the fastest, followed by MA-SGD.
Compared with GA-SGD, MA-SGD reduces the communication frequency from every batch to every epoch, which can be further reduced by ADMM.
Moreover, MA-SGD and ADMM can converge with fewer communication steps, in spite of reduced communication frequency.
We observe similar results when training {\bf SVM} on {\bf Higgs}: ADMM converges faster than GA-SGD and MA-SGD.

\vspace{0.3em}
\noindent{\bf (Results for MN)}
We have different observations when turning to training neural network models.
Figure~\ref{fig:dist_opt_mn_cifar10} presents the results of training {\bf MN} on {\bf Cifar10}.
First, we note that ADMM can only be used for optimizing convex objective functions and therefore is not suitable for training neural network models.
Comparing GA-SGD and MA-SGD, we observe that the convergence of MA-SGD is unstable, though it can reduce the communication cost.
On the other hand, GA-SGD can converge steadily and achieve a lower loss. As a result, in this case, we have no choice 
but to use GA-SGD.


\vspace{-0.5em}
\subsection{Communication Channels}
\label{medium}

\begin{quote}
\begin{small}
\em For many workloads, a pure FaaS architecture
can be competitive to the hybrid design with
a dedicated VM as parameter server, given the right
choice of the algorithm; A dedicated PS
can definitely help in principle, but its potential is
currently bounded by the communication between FaaS and IaaS.
\end{small}
\end{quote}

We evaluate the impact of communication channels.
We train {\bf LR}, {\bf MN}, and {\bf KM} using \system. 
{\bf LR} is optimized by ADMM, {\bf MN} is optimized by GA-SGD, and {\bf KM} is optimized by EM.
Table~\ref{tb:comm_medium} presents the settings and compares using disk-based S3 with other memory-based mediums.

\begin{table}[t!]
\scriptsize
\centering
\begin{tabular}{c c c c c c c}
\hline
\multirow{2}*{Workload} & \multicolumn{2}{c}{Memcached vs. S3}
& \multicolumn{2}{c}{DynamoDB vs. S3}
& \multicolumn{2}{c}{VM-PS vs. S3}  \\
\cline{2-7}
& cost & slowdown & cost & slowdown & cost & slowdown  \\
\hline
LR,Higgs,W=10 & 5 & 4.17 & 0.95 & 0.83 & 4.7 & 3.85  \\
LR,Higgs,W=50 & 4.5 & 3.70 & 0.92 & 0.81 & 4.47 & 3.70  \\
KMeans,Higgs,W=50,k=10 & 1.58 & 1.32 & 1.13 & 0.93 & 1.48 & 1.23  \\
KMeans,Higgs,W=50,k=1K & 1.43 & 1.19 & 1.03 & 0.90 & 1.52 & 1.27 \\
MobileNet,Cifar10,W=10 & 0.9 & 0.77 & N/A & N/A & 4.8 & 4.01  \\
MobileNet,Cifar10,W=50 & 0.89 & 0.75 & N/A & N/A & 4.85 & 4.03  \\
\hline
\end{tabular}
\caption{Comparison of S3, Memcached, DynamoDB, and VM-based parameter server. We present the slowdown and \emph{relative} costs of using different mediums w.r.t. using S3.
A relative cost larger than 1 means S3 is cheaper, whereas a slowdown larger than 1 means S3 is faster. DynamoDB cannot handle a large model such as MobileNet.
}
\vspace{-2em}
\label{tb:comm_medium}
\end{table}

\vspace{0.3em}
\noindent
{\bf (Pure FaaS Solutions)} 
We compare design choices including Memcached, S3,
Redis, and DynamoDB.

\begin{itemize}
    \item{\em Memcached vs. S3.}
    Memcached introduces a lower latency than S3, therefore one round of communication using Memcached is significantly faster than using S3.
    Furthermore, Memcached has a well-designed multi-threading architecture~\cite{lambda_redis_memcached}.
    As a result, its communication is faster than S3 over a large cluster with up to 50 workers, showing 7$\times$ and 7.7$\times$ improvements when training {\bf LR} and {\bf KM}.
    Nonetheless, the overall execution time of Memcached is actually longer than S3, because it takes more than two minutes to start Memcached whereas starting S3 is instant (as it is an ``always on'' service).
    When we turn to training {\bf MN} on {\bf Cifar10}, using Memcached becomes faster than using S3, since it takes much longer for {\bf MN} to converge.
    
    \item{\em Redis vs. Memcached.}
    According to our benchmark, Redis is similar to Memcached when training small ML models.
    However, when an ML model is large or is trained with a big cluster, Redis is inferior to Memcached since Redis lacks a similar high-performance multi-threading mechanism that underlies Memcached.
    
    \item{\em DynamoDB vs. S3.}
    Compared to S3, DynamoDB reduces the communication time by roughly 20\% when training LR on Higgs, though it remains significantly slower than IaaS if the startup time is not considered. 
    Nevertheless, DynamoDB only allows messages smaller than 400KB~\cite{aws_dynamodb_limits}, making it infeasible for many median models or large models (e.g., RCV1 and Cifar10).
    
\end{itemize}

\vspace{0.3em}
\noindent
{\bf (Hybrid Solutions)} \textsc{Cirrus}~\cite{cirrus} 
presents a hybrid design --- having a dedicated VM
to serve as parameter server and all FaaS 
workers communicate with this centralized PS.
This design definitely has its merit, in principle---giving the PS the ability of doing computation
can potentially save 2$\times$ communication compared with an FaaS communication channel via S3/Memcached.
However, we find that this hybrid design has several limitations, which limit the regime under which it outperforms a pure FaaS solution. 

When training {\bf LR} and {\bf KM}, VM-based PS performs similarly using Memcached or Redis, which are slower than S3 considering the start-up time. In this case,
a pure FaaS solution is competitive even without the 
dedicated VM. This is as expected---when the mode size is small and the runtime is relatively short, communication is not a significant bottleneck.

\begin{table}[!t]
\footnotesize
\centering
\begin{tabular}{c c c c c c}
\hline
\multirow{2}{*}{Lambda Type} & \multirow{2}{*}{EC2 Type} & Data Transmission & Model Update \\
 &  & gRPC / Thrift & gRPC / Thrift \\
\hline
1$\times$Lambda-3GB (1.8vCPU) & t2.2xlarge & 2.62s / 21.8s & 2.9s / 0.5s \\
1$\times$Lambda-1GB (0.6vCPU) & t2.2xlarge & 3.02s / 34.4s & 2.9s / 0.5s \\
1$\times$Lambda-3GB (1.8vCPU) & c5.4xlarge & 1.85s / 19.7s & 2.3s / 0.4s \\
1$\times$Lambda-1GB (0.6vCPU) & c5.4xlarge & 2.36s / 32s & 2.3s / 0.4s \\
10$\times$Lambda-3GB (1.8vCPU) & t2.2xlarge & 5.7s / 68.5s & 33s / 13s \\
10$\times$Lambda-1GB (0.6vCPU) & t2.2xlarge & 8.2s / 82s & 34s / 13s \\
10$\times$Lambda-3GB (1.8vCPU) & c5.4xlarge & 3.7s / 52s & 27s / 6s \\
10$\times$Lambda-1GB (0.6vCPU) & c5.4xlarge & 5.6s / 84s & 25s / 6s \\
\hline
\end{tabular}
\caption{Hybrid solution: Communication between Lambda and VM-based parameter server. 
Transferred data size is 75MB. The time is averaged over ten trials.
Transfer time includes time spent on {\em serialization/deserialization}.
In each pair, the {\em left} is result of {\em gRPC} and the right is result of {\em Thrift}.}
\label{tb:comm_faas_iaas}
\vspace{-1em}
\end{table}

When model is larger and workload is more communication-intensive, we would expect that the hybrid design performs significantly better. However, this is
not the case \textit{under the current infrastructure}.
To confirm our claim, we use two RPC frameworks (Thrift and gRPC), vary CPUs in Lambda (by varying memory size), use different EC2 types, and evaluate the communication between Lambda and EC2.
The results in Table~\ref{tb:comm_faas_iaas} reveal several constraints of communication between Lambda and VM-based parameter server:
(1) The communication speed from the PS is much slower than Lambda-to-EC2 bandwidth (up to 70MBps reported by~\cite{AnaATC18,serverless-ATC18}) and EC2-to-EC2 bandwidth (e.g., 10Gbps for c5.4xlarge).
Hybrid solution takes at least 1.85 seconds to transfer
75MB.
(2) Increasing the number of vCPUs can decrease the communication time by accelerating data serialization and deserialization.
But the serialization performance is eventually bounded by limited CPU resource of Lambda (up to 1.8 vCPU).
(3) Model update on parameter server is costly when the workload scales to a large cluster due to frequent locking operation of parameters.
As a result, HybridPS
is currently bounded not only by the maximal network bandwidth but also serialization/deserialization 
and model update.
\textit{However, if this
problem is fixed, we would expect that a hybrid design might be a perfect fit for FaaS-based deep learning. 
We will explore this in Section~\ref{sec:case_study}.}

We also study the impact of the number of parameter servers.
Intuitively, adding parameter servers can increase the bandwidth for model aggregation.
However, when we increase the number of parameter servers from 1 to 5 for the hybrid solution, we do not observe significant performance change.
As we explained above, the hybrid architecture is not bounded by the bandwidth; instead, the bottleneck is the serialization/deserialization operation in Lambda.
Therefore, adding parameter servers cannot solve this problem.



\vspace{-0.5em}
\subsection{Communication Patterns}

We use another model, called ResNet50 ({\bf RN}), in this study to introduce a larger model than {\bf MN}.
We train {\bf LR} on {\bf Higgs}, and train {\bf MN} and {\bf RN} on {\bf Cifar10}, using S3 as the external storage service.
Table~\ref{tb:anatomy_pattern_results} shows the time spent on communication by \texttt{AllReduce} and \texttt{ScatterReduce}.
We observe that using \texttt{ScatterReduce} is slightly slower than \texttt{AllReduce} when training {\bf LR}.
Here communication is not a bottleneck and \texttt{ScatterReduce} incurs extra overhead due to data partitioning.
On the other hand, the communication costs of \texttt{AllReduce} and \texttt{ScatterReduce} are roughly the same when training {\bf MN}.
\texttt{AllReduce} is 2$\times$ slower than \texttt{ScatterReduce} when training {\bf RN}, as communication becomes heavy and the single reducer (i.e., aggregator) in \texttt{AllReduce} becomes a bottleneck.

\begin{table}[t!]
\scriptsize
\centering
\begin{tabular}{c c c c c c}
\hline
Model \& Dataset & Model Size & AllReduce & ScatterReduce \\
\hline
LR,Higgs,W=50 & 224B & 9.2s & 9.8s \\
MobileNet,Cifar10,W=10 & 12MB & 3.3s & 3.1s \\
ResNet,Cifar10,W=10 & 89MB & 17.3s & 8.5s \\
\hline
\end{tabular}
\caption{Impact of different communication patterns.
}
\vspace{-1em}
\label{tb:anatomy_pattern_results}
\end{table}

\vspace{-0.5em}
\subsection{Synchronization Protocols}

\begin{figure}[!t]
    \centering
    \vspace{-1em}
    \includegraphics[width=0.8\columnwidth]{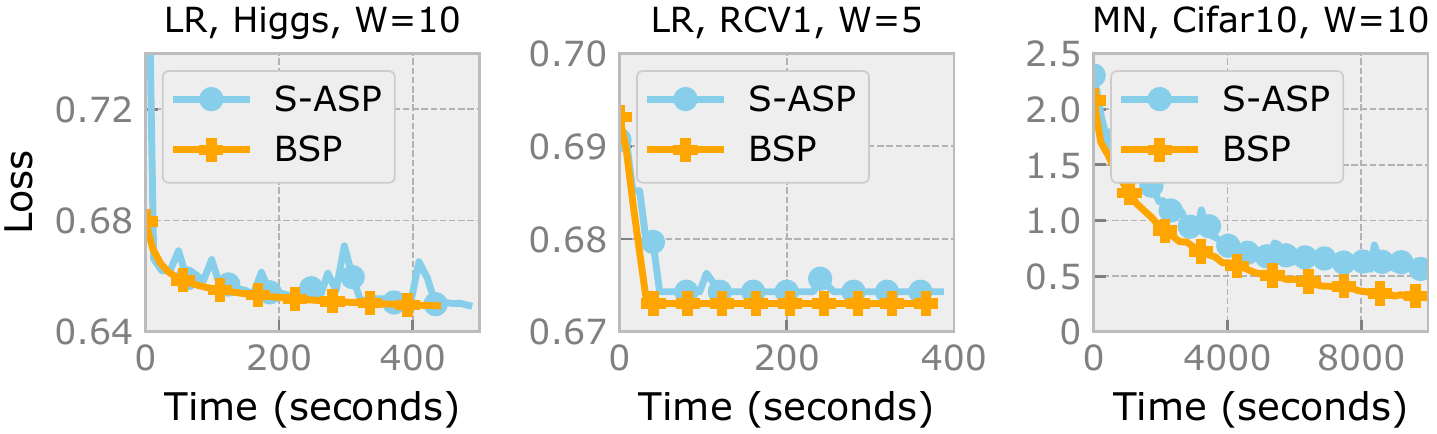}
    \vspace{-1em}
    \caption{Comparison of Synchronization Protocols.}
    \label{fig:sync_async}
\end{figure}

Finally, we study the impact of the two synchronization protocols: Synchronous and Asynchronous. Note that
the Asynchronous protocol here is different from 
ASP in traditional distributed learning.
In traditional distributed learning, ASP is implemented in the parameter-server architecture where there is an in-memory model replica that can be requested and updated by workers~\cite{dean2012large,ssp,HeteroSIGMOD}.
However, this ASP routine is challenging, if not infeasible, in FaaS infrastructure.
We thus follow SIREN~\cite{SIREN} to store a global model on S3 and let every FaaS instance rewrite it. This makes the impact of 
Asynchronous on convergence in our scenario 
more significant than that of ASP in distributed learning.
We use GA-SGD to train {\bf LR} on {\bf Higgs}, {\bf LR} on {\bf RCV1}, and {\bf MN} on {\bf Cifar10}, with Asynchronous or Synchronous enabled for the executors.
As suggested by previous work~\cite{async_sgd_delay}, we use a learning rate decaying with rate $1/\sqrt{T}$ for {\bf S-ASP} (our Asynchronous implementation) where $T$ denotes the number of epochs. Figure~\ref{fig:sync_async} presents the results.
We observe that Synchronous converges steadily whereas Asynchronous suffers from unstable convergence, although Asynchronous runs faster per iteration.
The convergence problem of Asynchronous is caused by the inconsistency between local model parameters. 
If stragglers exist, those faster executors may read stale model parameters from the stragglers.
Consequently, the benefit of system efficiency brought by Asynchronous is offset by its inferior statistical efficiency.

\vspace{-0.5em}
\section{FaaS vs. IaaS for ML Training}
\label{sec:serverless-versus-serverful}

We now compare FaaS and IaaS for ML training using \system.
Here we focus on the case of training a \emph{single model}. In this scenario, a user submits a training job over a dataset stored on S3; the system then (1) starts
the (FaaS or IaaS) infrastructure and (2) 
runs the job until it reaches the target loss. 


%

\vspace{-0.5em}
\subsection{Experiment Settings}
\label{seC:e2e_setup}

Our experimental protocols are as follows:

\vspace{0.3em}
\noindent
{\bf (Competing Systems)}
We compare \system, a pure FaaS-based implementation, with the following systems:
\begin{itemize}
\item{\em Distributed PyTorch.} We partition training data and run PyTorch 1.0 in parallel across multiple machines. 
We use all available CPU cores on each machine, if possible.
To manage a PyTorch cluster, we use StarCluster~\cite{starcluster}, a managing toolkit for EC2 clusters.
We use the \texttt{AllReduce} operator of Gloo, a collective communication library, for cross-machine communication, and we implement both mini-batch SGD and ADMM for training linear models.
\item{\em Distributed PyTorch on GPUs.} 
For deep learning models, we also consider GPU instances.
The other settings are the same as above.
\item{\em Angel.} Angel is an open-source ML system based on parameter servers~\cite{HeteroSIGMOD}. 
Angel works on top of the Hadoop ecosystem (e.g., HDFS, Yarn, etc.) and we use Angel 2.4.0 in our evaluation.
We chose Angel because it reports state-of-the-art performance on workloads similar to our evaluations.

\item{\em HybridPS.}
Following the hybrid architecture proposed by Cirrus~\cite{cirrus}, we implement a parameter server on a VM using gRPC, a cross-language RPC framework.
Lambda instances use a gRPC client to pull and push data to the parameter server.
We also implement the same SGD framework as in Cirrus.
\end{itemize}

\begin{table}[t!]
\scriptsize
\centering
\begin{tabular}{c c c c c c}
\hline
Model & Dataset & \# Workers  & Setting & Loss threshold \\
\hline
LR/SVM/KMeans & Higgs & 10 & $B$=10K,$k$=10 & 0.66/0.48/0.15 \\
LR/SVM & RCV1 & 5/5 & $B$=2K & 0.68/0.05 \\
KMeans & RCV1 & 50 & $k$=3 & 0.01 \\
LR/SVM/KMeans & YFCC100M & 100 & $B$=800,$k$=10 & 50 \\
MobileNet & Cifar10 & 10 & $B$=128 & 0.2 \\
ResNet50 & Cifar10 & 10 & $B$=32 & 0.4 \\
\hline
\end{tabular}
\caption{ML models, datasets, and experimental settings. $B$ means batch size, and $k$ means the number of clusters.}
\vspace{-1.5em}
\label{tb:ml_model_bench}
\end{table}

\noindent
{\bf (Datasets)}
In addition to {\bf Higgs}, {\bf RCV1} and {\bf Cifar10}, Figure~\ref{fig:e2e_dataset} presents two more datasets that are used for the current set of performance evaluations, {\bf YFCC100M} and {\bf Criteo}.
{\bf YFCC100M} (Yahoo Flickr Creative Commons 100 Million) is a computer vision~\cite{yfcc100m}, consisting of approximately 99.2 million photos and 0.8 million videos.
We use the YFCC100M-HNfc6~\cite{YFCC100M-HNfc6} in which each data point represents one image with several label tags and a feature vector of 4096 dimensions. We randomly sample 4 million data points, and convert this subset into a binary classification dataset by treating the ``animal'' tags as positive labels and the other tags as negative labels.
After this conversion, there are about 300K (out of 4M) positive data examples.
{\bf Criteo} is a famous dataset for click-through rate prediction hosted by Criteo and Kaggle.
Criteo is a high-dimensional sparse dataset with 1 million features.

\vspace{0.3em}
\noindent{\bf (ML Models)}
As shown in Table~\ref{tb:ml_model_bench}, we evaluate different ML models on different datasets, including {\bf LR}, {\bf SVM}, {\bf KM} and {\bf MN}.
We also consider a more complex deep learning model ResNet50. 
ResNet50 ({\bf RN}) is a famous neural network model that was the first to introduce identity mapping and shortcut connection.

\vspace{0.3em}
\noindent{\bf (EC2 Instances)}
We tune the optimal EC2 instance from the t2 family and the c5 family~\cite{ec2_pricing}.
To run PyTorch on GPUs, we tune the optimal GPU instances from the g3 family.
We use one c5.4xlarge EC2 instance as the parameter server in the hybrid architecture.

\vspace{0.3em}
\noindent{\bf (Protocols)}
We choose the optimal learning rate between 0.001 and 1.
We vary the number of
workers from 1 to 150.
Before running the competing systems, we partition the training data on S3.
We trigger Lambda functions after data is uploaded and Memcached is launched (if required).
We use one cache.t3.small Memcached node whose pricing is \$0.034 per hour.
Each ADMM round scans the training data ten times.
We stop training when the loss is below a threshold, as summarized in Table~\ref{tb:ml_model_bench}.

\begin{figure}[!t]
 \vspace{-1em}
 \centering
 \subfloat[\small LR, Higgs]{
    \label{fig:e2e_higgs_lr}
    \includegraphics[width=0.33\columnwidth]{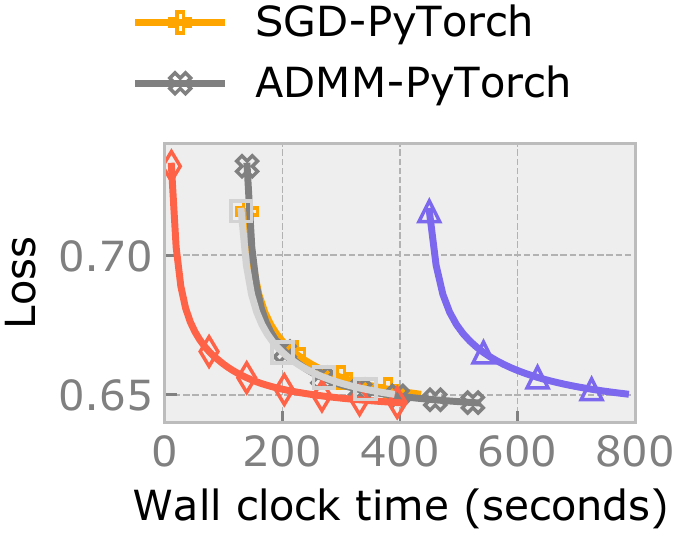}}
 \subfloat[\small SVM, Higgs]{
    \label{fig:e2e_higgs_svm}
    \includegraphics[width=0.33\columnwidth]{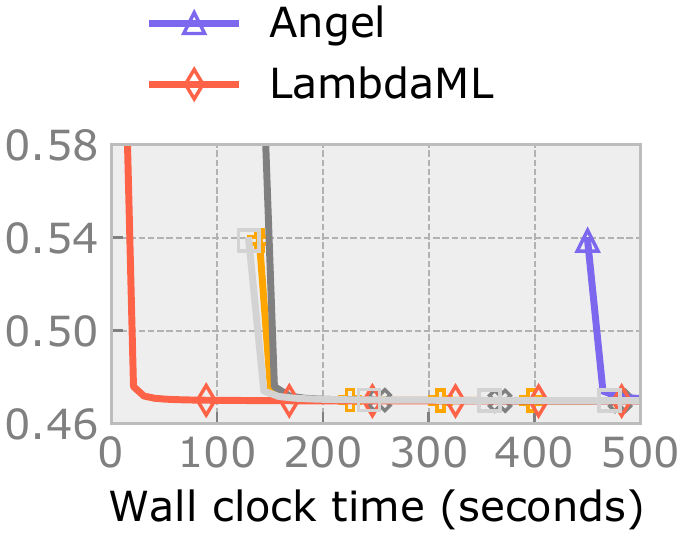}}
 \subfloat[\small KMeans, Higgs]{
    \label{fig:e2e_higgs_kmeans}
    \includegraphics[width=0.33\columnwidth]{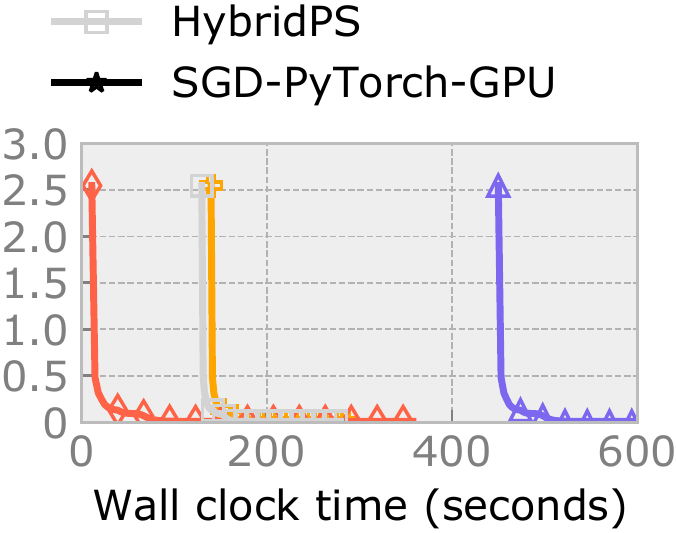}} \\
 \vspace{-1em}
 \subfloat[\small LR, RCV1]{
    \label{fig:e2e_rcv1_lr}
    \includegraphics[width=0.34\columnwidth]{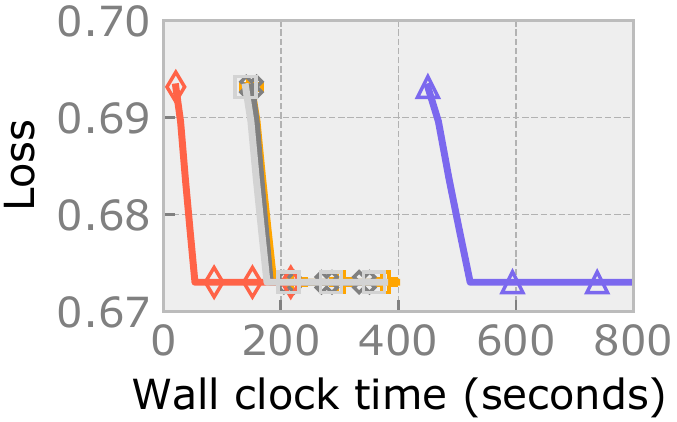}}
 \subfloat[\small SVM, RCV1]{
    \label{fig:e2e_rcv1_svm}
    \includegraphics[width=0.34\columnwidth]{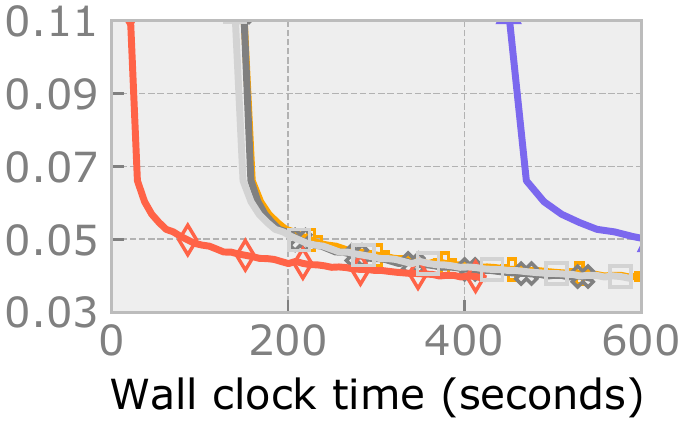}}
 \subfloat[\small KMeans, RCV1]{
    \label{fig:e2e_rcv1_kmeans}
    \includegraphics[width=0.32\columnwidth]{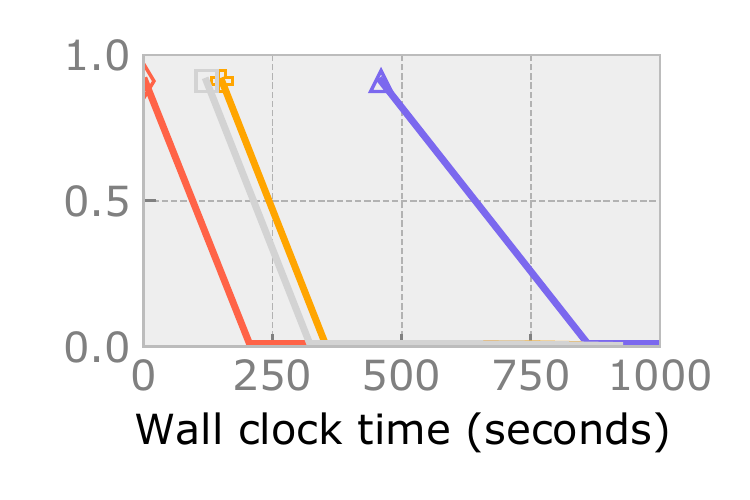}} \\
 \vspace{-1em}
 \subfloat[\small LR, YFCC100M]{
    \label{fig:e2e_yfcc_lr}
    \includegraphics[width=0.33\columnwidth]{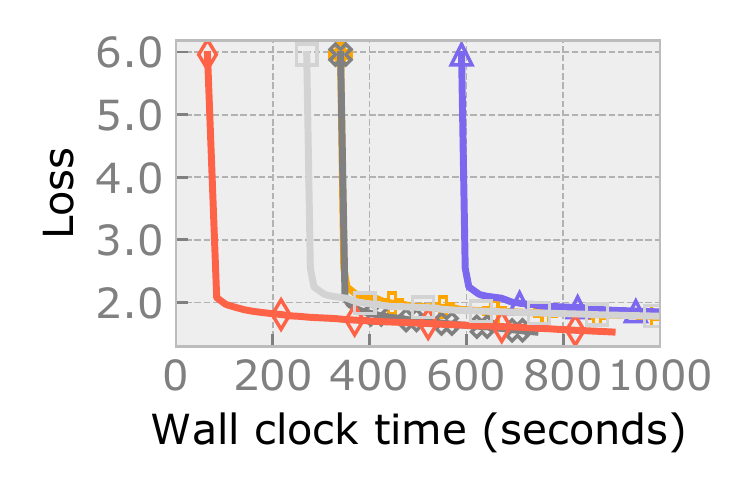}}
 \subfloat[\small SVM, YFCC100M]{
    \label{fig:e2e_yfcc_svm}
    \includegraphics[width=0.33\columnwidth]{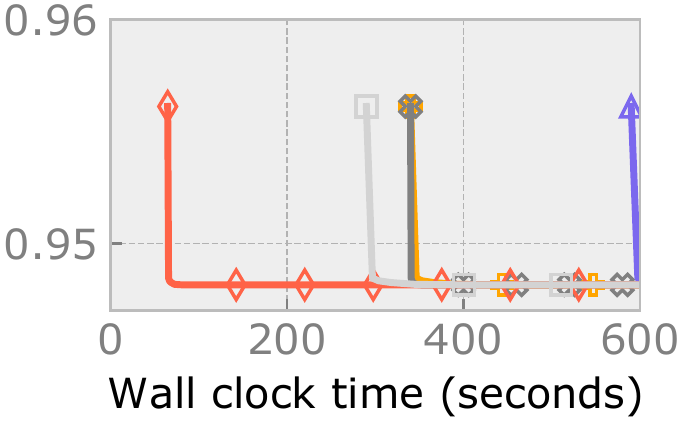}}
 \subfloat[\small KMeans, YFCC100M]{
    \label{fig:e2e_yfcc_kmeans}
    \includegraphics[width=0.33\columnwidth]{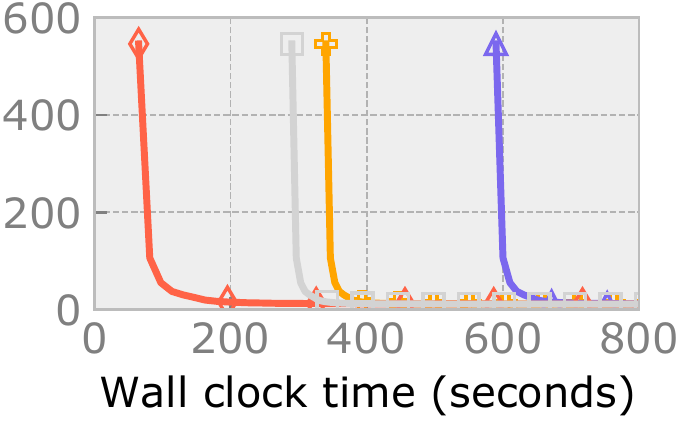}} \\
 \vspace{-1em}
 \subfloat[\small LR, Criteo]{
    \label{fig:e2e_criteo_lr}
    \includegraphics[width=0.33\columnwidth]{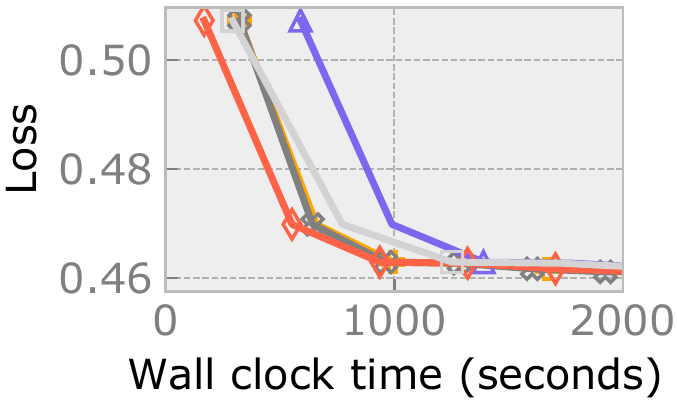}}
 \subfloat[\small MobileNet, Cifar10]{
    \label{fig:e2e_mobilenet}
    \includegraphics[width=0.33\columnwidth]{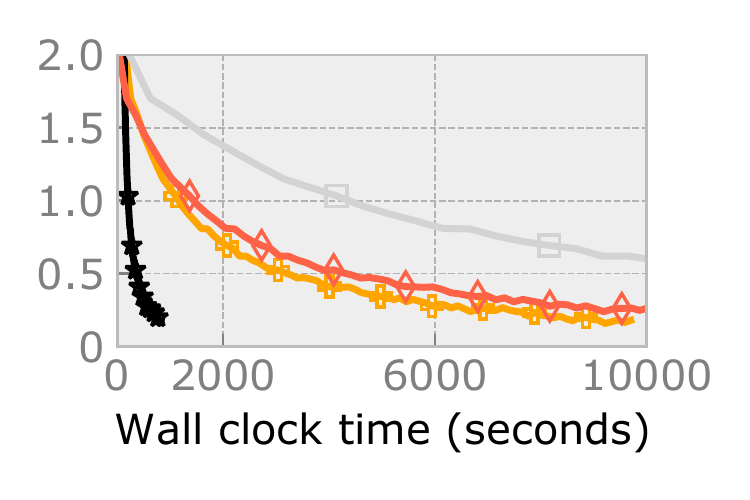}}
 \subfloat[\small ResNet50, Cifar10]{
    \label{fig:e2e_resnet}
    \includegraphics[width=0.33\columnwidth]{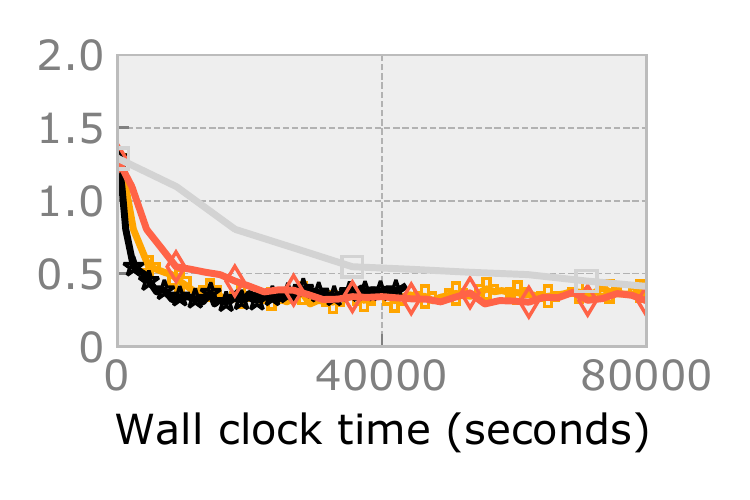}}
 \vspace{-1em}
 \caption{End-to-end comparison on various models.}
 \label{fig:e2e}
\end{figure}

\begin{figure}[!t]
  \centering
  \includegraphics[width=0.9\columnwidth]{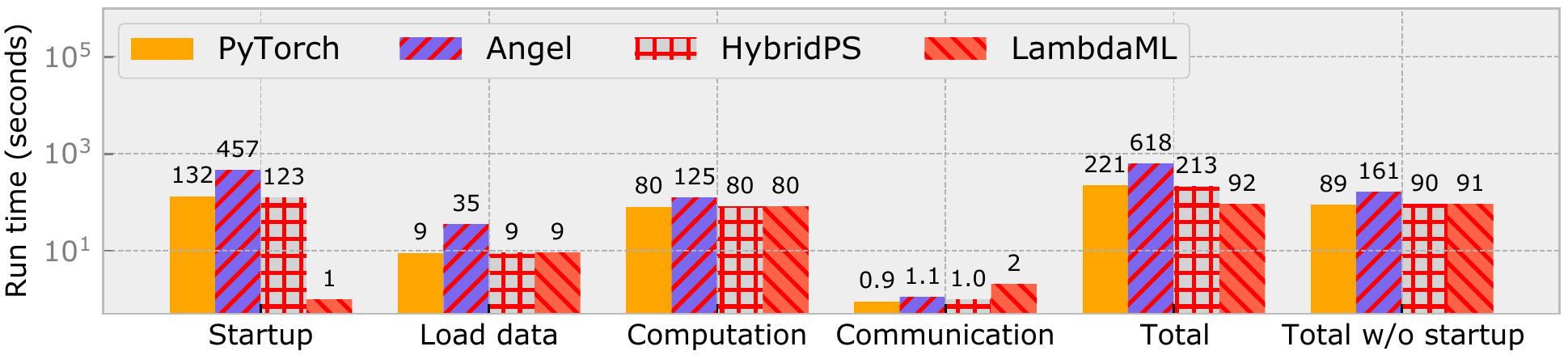}
  \vspace{-1em}
  \caption{Time breakdown (LR, Higgs, W = 10, 10 epochs).}
  \label{fig:e2e_time_breakdown}
\end{figure}

\vspace{-0.5em}
\subsubsection{``COST'' Sanity Check}

Before we report end-to-end experimental results,
we first report a sanity check as in COST~\cite{COST} to make sure all scaled-up solutions outperform a single-machine solution.
Taking Higgs and Cifar10 as examples, we store the datasets in a single machine and use a single EC2 instance to train the model and compare the performance with FaaS/IaaS.
For the Higgs dataset, using a single t2 instance (PyTorch) would converge in 960 seconds for LR trained by ADMM; while our FaaS (\system) and IaaS (distributed PyTorch) solutions, using 10 workers, converge in 107 and 98 seconds.
Similarly, on SVM/KMeans, FaaS and IaaS achieve 9.4/6.2 and 9.9/7.2 speedups using 10 workers; 
on Cifar10 and MobileNet, FaaS and IaaS achieve 4.8 and 6.7 speedups.

\vspace{-0.5em}
\subsection{Experimental Results}

We first present two micro-benchmarks using the same number of workers for FaaS and IaaS, and then discuss end-to-end results.

\vspace{0.5em}
\noindent
{\bf Algorithm Selection}. We first analyze the best algorithm
to use for each workload. We first run all competing
systems with the minimum
number of workers that can hold the dataset in memory.
We illustrate the convergence w.r.t wall-clock time  in 
Figure~\ref{fig:e2e}. 

\begin{quote}
\begin{small}
\em
In our design space, both FaaS and IaaS implementations
use the same algorithm (but not necessarily SGD) for
all workloads.
\end{small}
\end{quote}

\begin{enumerate}
    \item
    We first train {\bf LR}, {\bf SVM}, and {\bf KM} over Higgs, RCV1, and YFCC100M.
    Angel is the slowest as a result of slow start-up and computation.
    Running ADMM on PyTorch is slightly faster than SGD, verifying ADMM saves considerable communication while assuring convergence meanwhile.
    HybridPS outperforms PyTorch as it only needs to launch one VM and it is efficient in communication when the model is relatively small.
    \system achieves the fastest speed due to a swift start-up and the adoption of ADMM.
    To assess the performance of the baselines over high-dimensional features, we train models using the Criteo dataset.
    \system is still the fastest for LR (and the results on other models are similar) while the speed gap is smaller.
    This is unsurprising due to the high dimensionality of Criteo.
    
    \item
    For {\bf MN} and {\bf RN}, as analyzed above, data communication between Lambda and VM is bounded by the serialization overhead, and therefore the hybrid approach is slower than a pure FaaS approach with a large model.
    Distributed PyTorch is faster than \system because communication between VMs is faster than using ElastiCache in Lambda.
    PyTorch-GPU is the fastest as GPU can accelerate the training of deep learning models.
    The improvement of PyTorch-GPU on ResNet50 is smaller than on MobileNet because ResNet50 brings a heavier communication cost.
    For {\bf RN}, we also increase the batch size to 64, which incurs a larger memory footprint during the back-propagation training.
    FaaS encounters an out-of-memory error due to hitting the memory limit of 3GB, while 
    PyTorch can finish and achieve similar performance as using a batch size of 32.
    This demonstrates the limitation of the current FaaS infrastructure when training large models.
\end{enumerate}


\vspace{0.5em}
\noindent
{\bf Runtime Breakdown}.
To help understand the difference between FaaS and IaaS, 
Figure~\ref{fig:e2e_time_breakdown} presents a breakdown of time spent on executing 10 epochs, taking {\bf LR} on {\bf Higgs} as an example.

\begin{enumerate}
    \item \textbf{Start-up.} It takes more than 2 minutes to start a 10-node EC2 cluster, including the time spent on starting the VMs and starting the training job.
    The start-up of VMs also contains mounting shared volumes and configuring secure communication channels between VMs.
    The launching of a job also requires a master node dispensing scripts to all workers, meaning that it costs more time for a larger cluster.
    It takes even more time to start Angel, as it needs to first start dependent libraries such as HDFS and Yarn. The hybrid solution also needs to start and configure VMs, but it avoids the time spent on submitting job due to quick startup of FaaS functions. In contrast, \system took 1.3 seconds to start.
    \item \textbf{Data Loading and Computation.}
    In terms of data loading and computation, PyTorch, HybridPS, and \system spend similar amount of time because they all read datasets from S3 and use the same underlying training engine.
    Angel spends more time on data loading since it needs to load data from HDFS.
    Its computation is also slower due to inefficient matrix calculation library.
    \item \textbf{Communications.}
    Communication in \system is slower than in other baselines since \system uses S3 as the medium.
    
    \item {\bf Total Run Time.}
    In terms of the total run time, \system is the fastest when including the startup time.
    However, if the startup time is excluded, PyTorch outperforms \system since \system spends more time on communication.
    Especially, PyTorch does not incur start-up cost with reserved IaaS resources. 
\end{enumerate}

\begin{figure}[!t]
  \centering
  \subfloat{
    \label{fig:lr_higgs_scale_ondemand}
    \includegraphics[width=0.35\columnwidth]{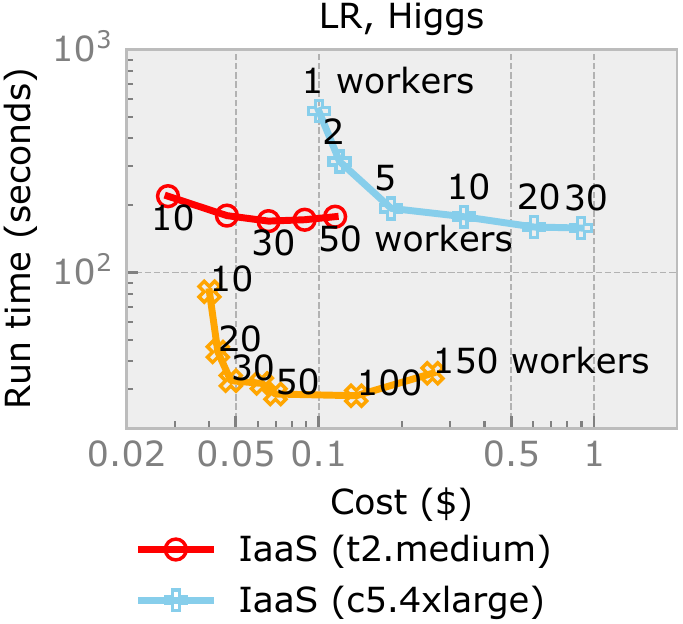}}
  \subfloat{
    \label{fig:mobile_cifar10_scale_ondemand}
    \includegraphics[width=0.35\columnwidth]{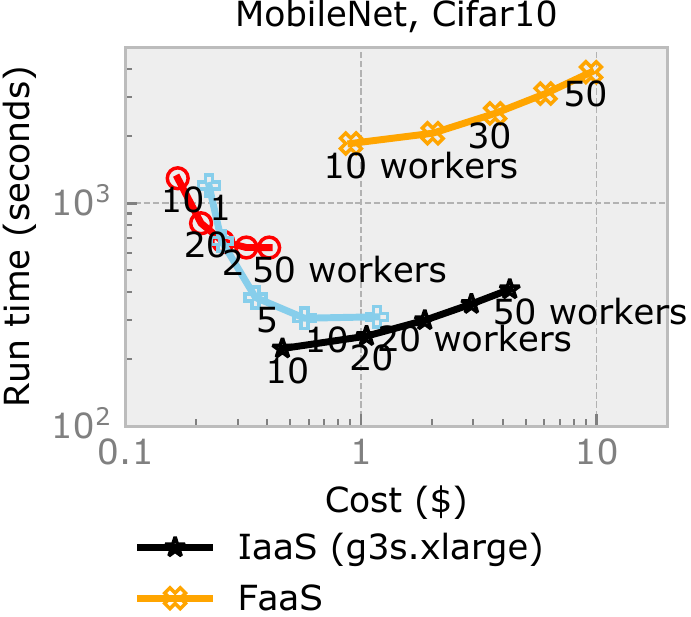}}
  \vspace{-1em}
  \caption{End-to-end comparison (w.r.t. \# workers).}
  \label{fig:cost_runtime_profile}
\end{figure}

\begin{figure}[!t]
  \centering
  \subfloat{
    \label{fig:lr_yfcc_scale_scatter}
    \includegraphics[width=0.35\columnwidth]{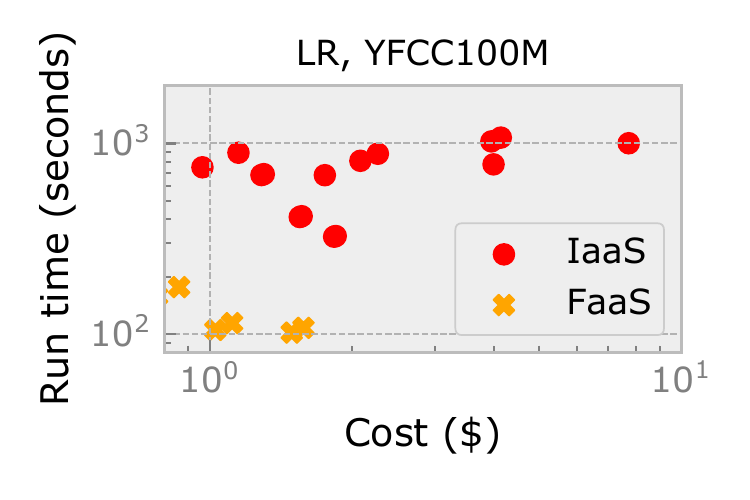}}
  \subfloat{
    \label{fig:svm_yfcc_scale_scatter}
    \includegraphics[width=0.35\columnwidth]{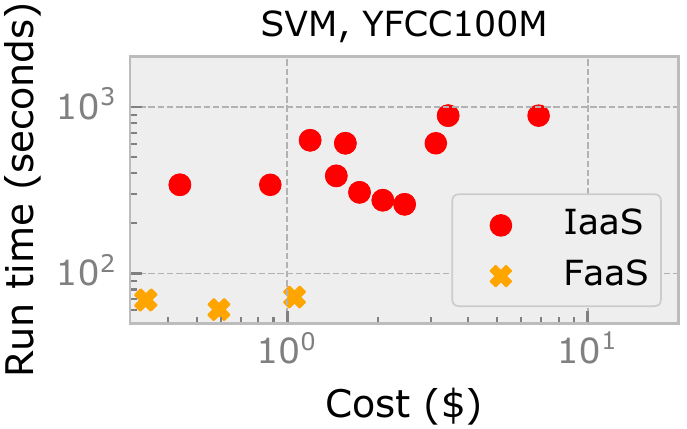}}\\
  \vspace{-1em}
  \subfloat{
    \label{fig:kmeans_yfcc_scale_scatter}
    \includegraphics[width=0.35\columnwidth]{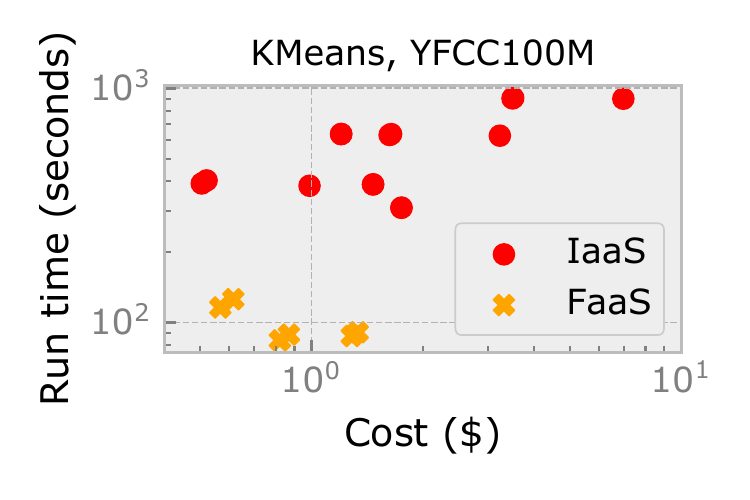}}
  \subfloat{
    \label{fig:mobilenet_cifar10_scale_scatter}
    \includegraphics[width=0.35\columnwidth]{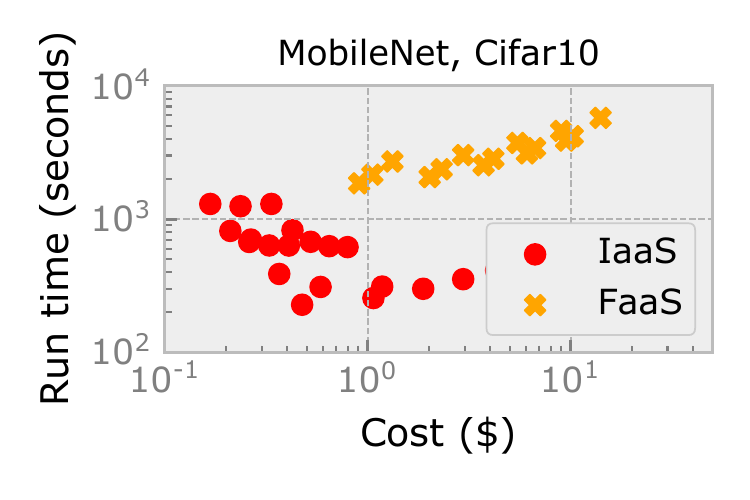}}
  \vspace{-1em}
  \caption{End-to-end comparison (w.r.t. various instance types and learning rates).}
  \label{fig:cost_runtime_scatter}
\end{figure}

\vspace{0.5em}
\noindent
{\bf End-to-end Comparison}. Comparing FaaS and IaaS by
forcing them to use the same number of workers is not
necessarily a fair comparison---an end-to-end comparison 
should also tune the optimal number of workers to use for each case.

\vspace{-0.3em}
\begin{quote}
\begin{small}
\em
\underline{For some communication-efficient workloads}, FaaS-based implementation can be significantly faster, but not significantly cheaper in dollar; \underline{on
other workloads}, FaaS-based implementation
can be significantly slower and more expensive.
\end{small}
\end{quote}

\vspace{-0.5em}
Figure~\ref{fig:cost_runtime_profile} illustrates two 
representative runtime vs. cost profiles. For models
that can take advantage of communication-efficient
algorithms (e.g., {\bf LR}, {\bf SVM}, {\bf Kmeans}), adding workers initially
makes both FaaS and IaaS systems faster, and then flattened
(e.g., FaaS at 100 workers). Different systems
plateaued at different runtime levels, illustrating the
difference in its start-up time and communication cost. 
On the other hand, the more workers we add, the more
costly the execution is. For models that cannot 
take advantage of communication-efficient
algorithms (e.g., {\bf MN}, {\bf RN}), the FaaS system flattened earlier, illustrating the hardness of scale-up.

In Figure~\ref{fig:cost_runtime_scatter}, we plot
for different configurations such as 
learning rate and instance type (GPU instance for {\bf MN}).
In addition to G3 GPU instances (NVIDIA M60), we also consider a G4 GPU instance (NVIDIA T4) for {\bf MN}.
The red points refer to IaaS systems, and the orange points refer to FaaS systems.
Note that there are more red points than orange points because we need to tune different instance types for IaaS.
For {\bf LR} and {\bf SVM}, there is an 
FaaS configuration that 
is faster than all IaaS configurations in terms
of runtime; however, they are not significantly cheaper, mirroring
the result similar to Lambada~\cite{lambada}
and Starling~\cite{starling}.
For
{\bf KMeans}, a user minimizing for cost
would use IaaS while FaaS implementations
are significantly faster if the user
optimizes for runtime. For {\bf MN}, 
the opposite is true --- there exists an 
IaaS configuration that outperforms \textit{all}
FaaS configurations in \textit{both} runtime and cost ---
using T4 GPUs is 8$\times$ faster and 9.5$\times$ cheaper than the best FaaS execution (15\% faster and 30\% cheaper than M60).

\begin{table}
\scriptsize
\centering
\begin{tabular}{c c c c c}
\hline
Workload & Run time  & Test accuracy & Cost\\
\hline
FaaS (LR,Higgs,W=10) & 96 & 62.2\% & 0.47 \\
IaaS (LR,Higgs,W=10) & 233 & 62.1\% & 0.31 \\
FaaS (MobileNet,Cifar10,W=10) & 1712 & 80.45\% & 8.37 \\
IaaS (MobileNet,Cifar10,W=10) & 1350 & 80.52\% & 1.74 \\
\hline
\end{tabular}
\caption{ML Pipeline (time is in seconds and cost is in \$).}
\vspace{-1.5em}
\label{tb:pipeline}
\end{table}

\vspace{0.5em}
\noindent
{\bf Pipeline Comparison.}
A real-world training workload of an ML model is typically a pipeline, consisting of data preprocessing, model training, and hyperparameter tuning.
To assess the performance of IaaS- and FaaS-based systems on ML pipelines, we construct a pipeline using Higgs and Cifar10:
(1) normalize original features to $[-1,1]$, and (2) grid-search the learning rate in the range $[0.01,0.1]$ with an increment of 0.01.
We perform preprocessing using a single job with 10 workers,
and parallelize hyperparameter tuning using multiple jobs (each job with 10 workers and 10 epochs).
For IaaS, we use ten t2.medium instances as workers.
For FaaS, we (1) use a serverless job to perform the preprocessing and store the transformed dataset to S3, and then (2) trigger one serverless job for each hyperparameter, using S3 as the communication medium.
The other settings are the same as in Section~\ref{sec:eval_lambda_setting}.
As Table~\ref{tb:pipeline} shows, we observe similar results as in the end-to-end comparison.
FaaS is faster than IaaS for LR, however,
is not cheaper. For MobileNet, 
IaaS is significantly faster and cheaper.

\vspace{-0.5em}
\subsection{Analytical Model} \label{sec:analytical}

Based on the empirical observations, we now develop an analytical model that captures the cost/performance tradeoff between different configuration points in the design space covered in Section~\ref{sec:serverless}.

Given an ML task, for which
the dataset size is $s$ MB and the model size
is $m$ MB, let the start-up time of $w$
FaaS (resp. IaaS) workers be $t^{F}(w)$ 
(resp. $t^{I}(w)$), the \textit{bandwidth} of 
S3, EBS, network, and ElastiCache be $B_{S3}$,
$B_{EBS}$, $B_{n}$, $B_{EC}$, the \textit{latency} of S3, EBS, network, and ElastiCache be $L_{S3}$,
$L_{EBS}$, $L_{n}$, $L_{EC}$.
Assuming that the algorithm used by FaaS (resp. IaaS) 
requires $R^{F}$ (resp. $R^{I}$) epochs to converge with 
one single worker, we use $f^{F}(w)$ (resp. $f^{I}(w)$)
to denote the ``scaling factor'' of convergence
which means that using $w$ workers will
lead to $f^{F}(w)$ times more epochs.
Let $C^{F}$ (resp. $C^{I})$ be the time
that a single worker needs for computation 
of a single epoch.
With $w$ workers, the execution time of FaaS and IaaS
can be modeled as follows (to model the cost in dollar 
we can simply multiply the unit cost per second):

\begin{footnotesize}
\vspace{-1em}
\[
FaaS(w):= 
\underbrace{\textcolor{darkgreen}{t^{F}(w)} + \frac{s}{B_{S3}}}_\text{start up \& loading} +
\overbrace{R^{F} f^{F}(w)}^\text{convergence} \times 
\underbrace{\left(\textcolor{red}{ (3w-2)(\frac{m/w}{B_{S3/EC}} + L_{S3/EC}) }\right.}_\text{communication} + \overbrace{\left.\frac{C^F}{w}\right)}^\text{computation}, 
\]
\vspace{-1.5em}
\end{footnotesize}

\begin{footnotesize}
\vspace{-1em}
\[
IaaS(w) :=
\underbrace{\textcolor{red}{ t^{I}(w)} + \frac{s}{B_{S3}} }_\text{start up \& loading} +
\overbrace{R^{I} f^{I}(w)}^\text{convergence} \times 
\underbrace{\left(\textcolor{darkgreen}{ (2w-2) (\frac{m/w}{B_{n}} + L_{n}) }\right.}_\text{communication} + \overbrace{\left.\frac{C^I}{w}\right)}^\text{computation}, 
\]
\vspace{-0.8em}
\end{footnotesize}

where the color-coded terms represent the ``built-in''
advantages of FaaS/IaaS (green means
holding advantage) --- FaaS incurs smaller start-up
overhead, while IaaS incurs smaller communication overhead
because of its flexible mechanism
and higher bandwidth.
The difference in the constant,
i.e., $(3w - 2)$ and $(2w - 2)$,
is caused by the fact that
FaaS can only communicate via storage services that do not
have a computation capacity (Section~\ref{sec:design_communication_channel}). The latency term $L_{S3/EC}$ and $L_n$ could dominate for smaller messages.

\vspace{0.5em}
\noindent
\underline{\textit{When will FaaS outperform IaaS?}}
From the above analytical 
model it is clear to see the regime under which FaaS 
can hold an edge. This regime is defined by two properties: (1) \textit{scalability}--the FaaS algorithm 
needs to scale well, i.e., a small $f^F(w)$ (thus a
small $\frac{f^F(w)}{w}$ for large $w$), 
such that the overall cost is not
dominated by computation; and (2) \textit{communication
efficiency}--the FaaS algorithm needs to converge 
with a small number of rounds, i.e., a small $R^F$.

\begin{table}[t!]
\scriptsize
\centering
\begin{tabular}{c c c c c c}
\hline
\textbf{Symbol} & \textbf{Configurations} & \textbf{Values} \\
\hline
$t^{F}(w)$ & $w$=10,50,100,200 & (1.2$\pm$0.1)s,(11$\pm$1)s,(18$\pm$1)s,(35$\pm$3)s \\
$t^{I}(w)$ & $w$=10,50,100,200 & (132$\pm$6)s,(160$\pm$5)s,(292$\pm$8)s,(606$\pm$12)s \\
$B_{S3}$ & Amazon S3 & (65$\pm$7)MB/s \\
$B_{EBS}$ & gp2 & (1950$\pm$50)MB/s \\
$B_{n}$ & t2.medium to t2.medium & (120$\pm$6)MB/s \\
$B_{n}$ & c5.large to c5.large & (225$\pm$8)MB/s \\
$B_{EC}$ & cache.t3.medium & (630$\pm$25)MB/s \\
$B_{EC}$ & cache.m5.large & (1260$\pm$35)MB/s \\
$L_{S3}$ & Amazon S3 & (8$\pm$2)$\times 10^{-2}$s \\
$L_{EBS}$ & gp2 & (3$\pm$0.5)$\times 10^{-5}$s \\
$L_{n}$ & t2.medium to t2.medium & (5$\pm$1)$\times 10^{-4}$s\\
$L_{n}$ & c5.large to c5.large & (1.5$\pm$0.2)$\times 10^{-4}$s\\
$L_{EC}$ & cache.t3.medium & (1$\pm$0.2)$\times 10^{-2}$s \\
\hline
\end{tabular}
\caption{Constants for the analytical model.}
\vspace{-2em}
\label{tb:analytical_constans}
\end{table}

\begin{figure}
    \centering
    \subfloat[Analytical Model vs. Actual Runtime.]{
    \includegraphics[width=0.35\columnwidth]{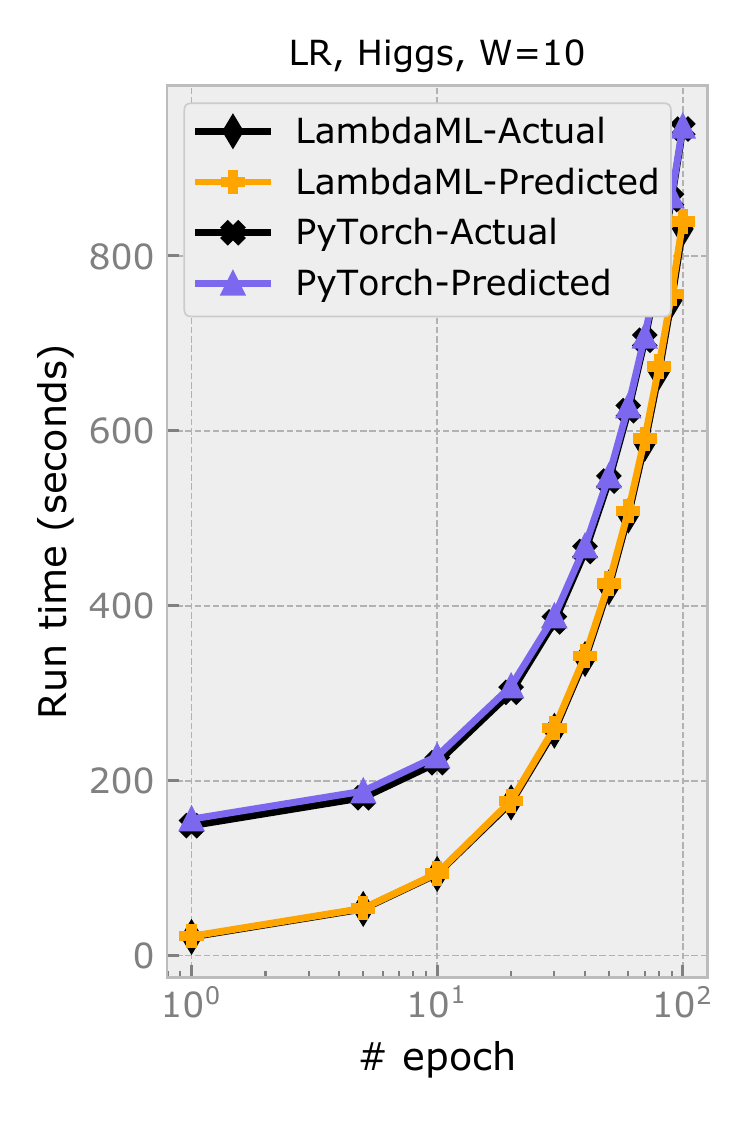}}
    \hspace{0.5em}
    \subfloat[Sampling-based Estimator plus. Analytical Model.]{
    \includegraphics[width=0.55\columnwidth]{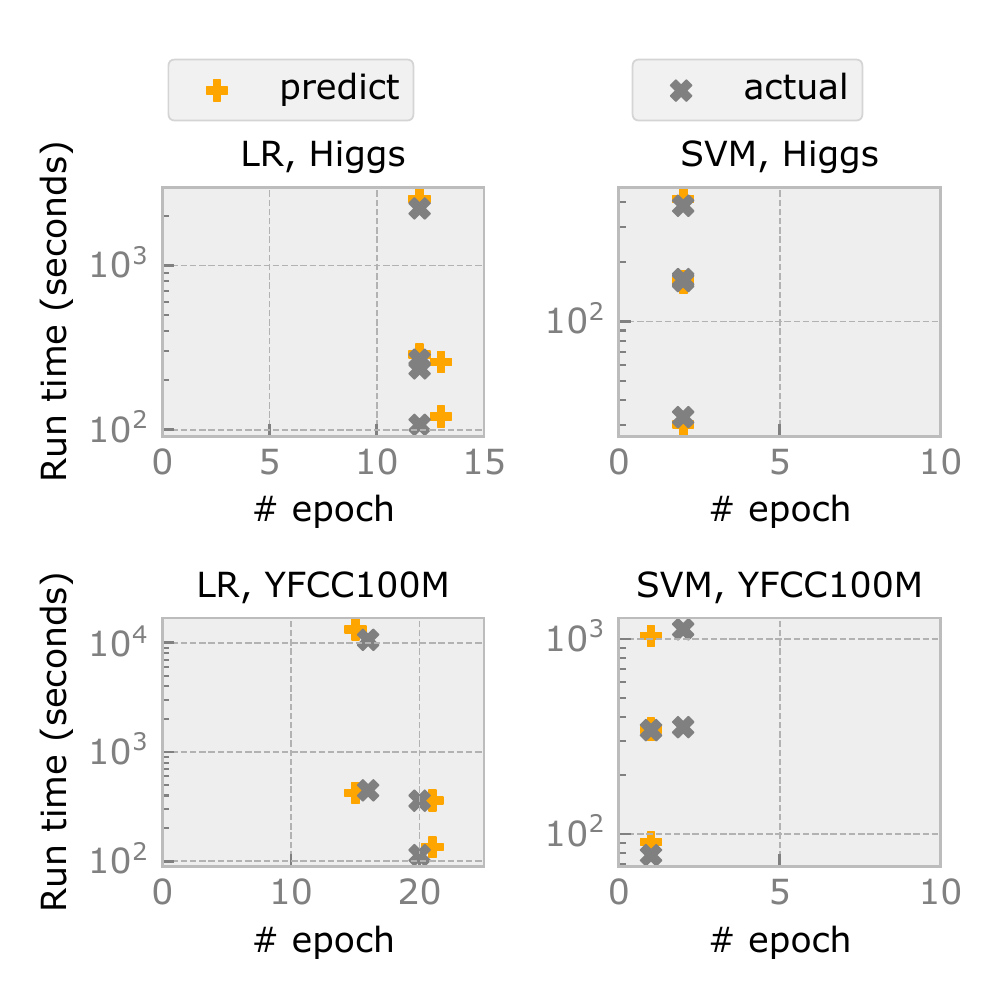}}
    \vspace{-1em}
    \caption{Evaluation of Analytical Model.}
    \label{fig:analytical_accuracy}
\end{figure}

\vspace{0.5em}
\noindent
{\bf (Validation)}
We provide an empirical validation of
this analytical model. First,
we show that \textit{given the right constant,
this model correctly reflects the runtime
performance of FaaS-based and IaaS-based systems.}
We train a logistic regression model 
on Higgs with ten workers (the results on other models and datasets are similar) and show the analytical model vs. the actual 
runtime in Figure~\ref{fig:analytical_accuracy}(a).
Across a range of fixed number of epochs (from 1 to 100), 
and using the constant in Table~\ref{tb:analytical_constans},
we see that the analytical model approximates the actual runtime reasonably well.

The goal of our analytical model is 
to understand the fundamental tradeoff governing the runtime performance, instead of serving as a predictive 
model. To use it as a predictive model one has to 
estimate the number of epochs that each 
algorithm needs. This problem has been the focus 
of many previous works (e.g., \cite{cost_optimizer_sigmod})
and is thus orthogonal to this paper. Nevertheless,
we implement the sampling-based estimator 
in \cite{cost_optimizer_sigmod} and use 10\% of training data
to estimate the number of epochs needed.
Then, the estimated epoch numbers and unit runtime, together with the constant in Table~\ref{tb:analytical_constans}, are used to predict the end-to-end runtime.
We choose four workloads (LR/SVM \& Higgs/YFCC100M) and train them with two optimization algorithms (SGD and ADMM) in both FaaS (\system) and IaaS (distributed PyTorch).
Figure~\ref{fig:analytical_accuracy}(b) shows that
this simple estimator can estimate the number of epochs well for both SGD and ADMM, and
the analytical model can also estimate the runtime accurately.
It is interesting future work
to develop a full-fledged predictive model for FaaS and IaaS by combining the insights obtained from this paper
and \cite{cost_optimizer_sigmod}.

\vspace{-0.5em}
\subsubsection{Case Studies}
\label{sec:case_study}

We can further use this analytical model to explore alternative configurations in future infrastructures


\begin{figure}
  \centering
  \includegraphics[width=0.35\textwidth]{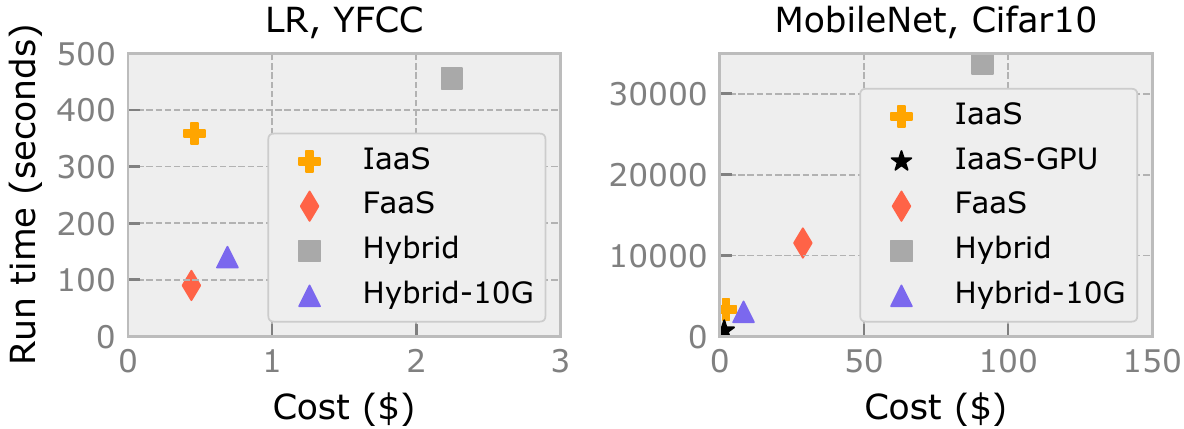}
  \vspace{-1em}
  \caption{Simulation: Faster FaaS-IaaS Communication.}
  \label{fig:taas_ondemand_time_faster_hybrid}
\end{figure}

\vspace{0.5em}
{\em Q1: What if Lambda-to-VM communication becomes faster (and support GPUs)?}
As we previously analyzed, the performance of HybridPS is bounded by the communication speed between FaaS and IaaS.
How would accelerating the FaaS-IaaS communication change our tradeoff?
This is possible in the future by having higher FaaS-to-IaaS bandwidth, faster RPC frameworks, or more CPU resources in FaaS.
To simulate this, we assume bandwidth between FaaS and IaaS can be fully utilized and change the bandwidth to 10GBps in our analytical model.
As shown in Figure~\ref{fig:taas_ondemand_time_faster_hybrid}, the performance of HybridPS could be significantly improved.
When training {\bf LR} over {\bf YFCC100M}, HybridPS-10GBps is worse than FaaS since FaaS saves the start-up time of one VM and uses ADMM instead of SGD.
When training {\bf MN} over {\bf Cifar10}, HybridPS-10GBps would be about 10\% faster than IaaS; however it is still slower than IaaS-GPU.

If future FaaS further supports GPUs and offers similar pricing compared with comparable IaaS infrastructure  --- \$0.75/hour for g3s.xlarge ---
HybridPS-10GBps would be 18\% cheaper than IaaS. This would make FaaS a promising platform for training deep neural networks; otherwise, under the current
pricing model, IaaS is still more cost-efficient even compared with HybridPS-10GBps.

%
%

\begin{figure}
  \centering
  \includegraphics[width=0.35\textwidth]{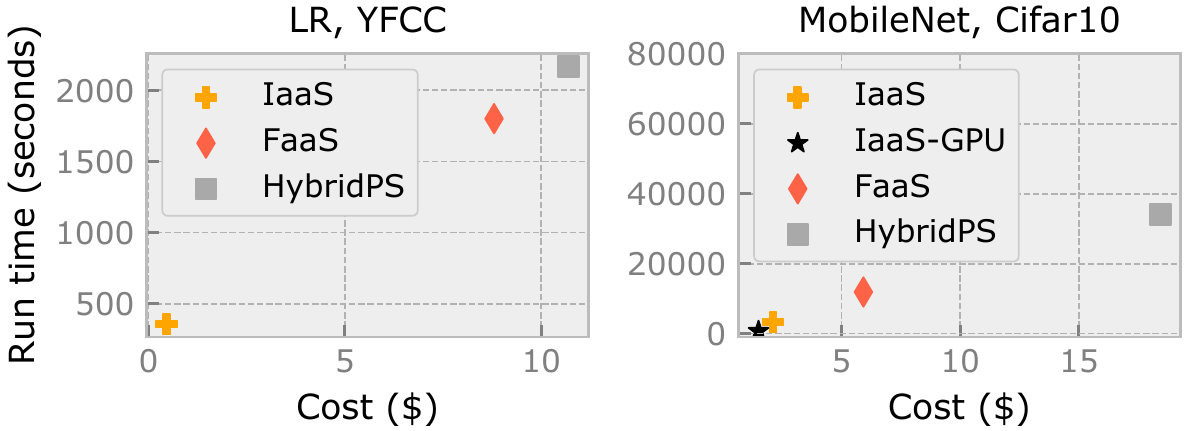}
  \vspace{-1em}
  \caption{Simulation: Hot Data.}
  \label{fig:taas_ondemand_time_hot_data}
\end{figure}

\vspace{0.5em}
{\em Q2: What if the data is hot?} 
Throughout this work, we assumed that both FaaS-based and IaaS-based implementations read data from remote disk such as S3. What if the data is ``hot'' and has already been stored in
the VM? To understand this, suppose that the {\bf YFCC100M}
dataset is stored in a powerful VM (m5a.12xlarge). IaaS-based system (t2.medium), FaaS-based system, and HybridPS all read the dataset from the VM.
As shown in Figure~\ref{fig:taas_ondemand_time_hot_data}, IaaS significantly outperforms FaaS and HybridPS, due to the slow communication when FaaS reads hot data. This is consistent with observations by
Hellerstein et al.~\cite{lambada,starling,serverless-CIDR19} for ML training, and resonates observations by previous work on non-ML workloads~\cite{lambada,starling}.

\vspace{0.5em}
{\em Q3: What about multi-tenancy?} 
When the system is serving multiple users/jobs,
FaaS-based solution can potentially provide 
a more flexible provisioning strategy: 
start an FaaS run for each job on-demand. Given 
the short start-up time of FaaS-based solution,
this might provide a benefit over both reserved 
VM solutions and on-demand VM solutions, especially 
for ``peaky'' workloads. This observation has 
been reported for other non-ML workloads~\cite{lambada}
and we expect something similar for ML training.
Given the space limitations, we leave this 
aspect to future work.

\vspace{-0.5em}
\section{Related Work}
\label{sec:related-work}

\noindent{\bf (Distributed ML)}
Data parallelism is a common strategy used by distributed ML systems, which partitions and distributes data evenly across workers.
Each worker executes the training algorithm over its local partition, and synchronizes with other workers from time to time.
A typical implementation of data parallelism is parameter server~\cite{dean2012large, ssp, li2014communication, dai2015high, dimboost,dynamic_ps,tensorflow}.
Another popular implementation is message passing interface (MPI)~\cite{gropp1999using}, e.g., the \texttt{AllReduce} MPI primitive leveraged by MLlib~\cite{mllib}, XGBoost~\cite{xgboost}, PyTorch~\cite{dist_pytorch}, etc~\cite{MLBase}.
We have also used data parallelism to implement \system.
Other research topics in distributed ML include compression~\cite{qsgd,terngrad,zipml,skcompress,sketchml,ATOMO,error_compen_qsgd,convergence_sparsified},
decentralization~\cite{compress_decentral,decentral_nips,d2_decentral,convergence_decentral_siam,dual_avg_decentral_ICML,COLA_decentral_linear,decentral_compress_Jaggi},
synchronization~\cite{async_noise_Duchi,comm_sgd_Duchi,delayed_sgd_Duchi,local_sgd_Jaggi,cocoa_Jaggi,cocoa_plus_Jaggi,ada_comm,gaia_geo_ML,async_prox_ICML},
straggler~\cite{grad_coding,hogwild,straggler_encoding,async_compensate,Zeno,byzantine_sgd},
data partition~\cite{ColumnML,vertical_gbdt,singa,column_vs_row,Yggdrasil}, etc.

\vspace{0.5em}
\noindent{\bf (Serverless Data Processing)}
Cloud service providers have introduced their serverless platforms, such as AWS Lambda~\cite{lambda}, Google Could Functions~\cite{google_function}, and Azure Functions~\cite{azure}.
Quite a few studies have been devoted to leveraging these serverless platforms for large-scale data processing.
For example, Locus~\cite{ShuffleServerless} explores the trade-off of using fast and slow storage mediums when shuffling data under serverless architectures.
Numpywren~\cite{Numpywren} is an \emph{elastic} linear algebra library on top of a serverless architecture.
Lambada~\cite{lambada} designs an efficient invocation approach for TB-scale data analytics.
Starling~\cite{starling} proposes a serverless query execution engine. 

\vspace{0.5em}
\noindent{\bf (Serverless ML)}
Building ML systems on top of serverless infrastructures has emerged as a new research area.
Since ML model inference is a straightforward use case of serverless computing~\cite{Stratum,ServeServerless}, the focus of recent research effort has been on ML model training.
For instance, Cirrus~\cite{cirrus} is a serverless framework that supports end-to-end ML workflows. 
In~\cite{ServerlessNN}, the authors studied training neural networks using AWS Lambda. 
SIREN~\cite{SIREN} proposes an asynchronous distributed ML framework based on AWS Lambda. 
Hellerstein et al.~\cite{serverless-CIDR19} show $21\times$ to $127\times$
performance gap, with FaaS lagging behind 
IaaS because of the overhead of data loading and the
limited
computation power. 
Despite these
early explorations, it remains challenging 
for a practitioner to
reach a firm conclusion about the relative 
performance of FaaS and IaaS for ML Training. 
On the other hand, Fonseca et al.~\cite{cirrus}
in their \textsc{Cirrus} system, Gupta et al.~\cite{Oversketched}
in their \textsc{OverSketched Newton} algorithm, and
Wang et al.~\cite{SIREN} in their \textsc{SIREN} system, depict a
more promising picture in which FaaS is
$2\times$ to $100\times$ faster than IaaS on a range 
of workloads. 
The goal of this work is to provide 
a systematic, empirical study.

\section{Conclusion}
\label{sec:conclusion}
We conducted a systematic study regarding 
the tradeoff between FaaS-based and IaaS-based 
systems for training ML models.
We started by an anatomy of the design space that covers the optimization algorithm, the communication channel, the communication pattern, and the synchronization protocol, which had yet been explored by previous work. We then implemented LambdaML, a prototype system of FaaS-based training on Amazon Lambda, following which we 
systematically depicted the tradeoff space and 
identified cases where FaaS holds an edge.
Our results indicate that ML training pays off in serverless infrastructures only for models with efficient (i.e., reduced) communication and that quickly converge. In general, FaaS can be much faster but it is never significantly cheaper than IaaS.  

\section*{Acknowledgment}

{\footnotesize
We are appreciative to all
anonymous reviewers at 
VLDB'21 and 
SIGMOD'21, who provide insightful feedback that makes this paper much stronger. We 
also appreciate the help from 
our anonymous shepherd and area chair at SIGMOD'21 to their 
shepherding.}

{\footnotesize
CZ and the DS3Lab gratefully acknowledge the support from the Swiss National Science Foundation (Project Number 200021\_184628), Innosuisse/SNF BRIDGE Discovery (Project Number 40B2-0\_187132), European Union Horizon 2020 Research and Innovation Programme (DAPHNE, 957407), Botnar Research Centre for Child Health, Swiss Data Science Center, Alibaba, Cisco, eBay, Google Focused Research Awards, Oracle Labs, Swisscom, Zurich Insurance, Chinese Scholarship Council, and the Department of Computer Science at ETH Zurich.
}

\clearpage

\balance
\bibliographystyle{ACM-Reference-Format}
\bibliography{main}

\end{document}